\documentclass[
 superscriptaddress,
 amsmath,amssymb,
 aps,
 prb,
 reprint,
 longbibliography,
 nofootinbib
]{revtex4-2}

\usepackage{graphicx}
\usepackage{dcolumn}
\usepackage{bm}
\usepackage[dvipsnames]{xcolor}
\usepackage{tikz}
\usepackage{xfrac}
\usepackage{blindtext}
\usepackage{times}
\usepackage{multirow}
\usepackage{dsfont}

\usepackage{color}
\usepackage[colorlinks=true, urlcolor=blue, linkcolor=blue, citecolor=blue, pdftex]{hyperref}

\usepackage[normalem]{ulem}

\definecolor{darkblue}{HTML}{004D6B}
\definecolor{darkred}{HTML}{8c1515}
\definecolor{darkgreen}{HTML}{006400}
\usepackage{hyperref}
\hypersetup{
	colorlinks=true,
	urlcolor=darkred,
	citecolor=darkblue,
	linkcolor=darkred,
	breaklinks
}

\usepackage[caption=false]{subfig}

\graphicspath{{Figures/}}

\newcommand{\pdagger}{{\phantom{\dagger}}}
\newcommand{\btau}{{\boldsymbol{\tau}}}
\newcommand{\btaup}{{\boldsymbol{\tau}^\perp}}
\newcommand{\bsigma}{{\boldsymbol{\sigma}}}
\newcommand{\svsym}{SU(2)${}_{\mathrm{spin}}\otimes$U(1)${}_{\mathrm{valley}}$}
\newcommand{\mev}{\ \mathrm{meV}}

\begin{document}

\title{Spin-valley magnetism on the triangular moir\'e lattice with SU(4)~breaking interactions}

\author{Lasse Gresista}
\email{gresista@thp.uni-koeln.de}
\affiliation{Institute for Theoretical Physics, University of Cologne, 50937 Cologne, Germany}
\author{Dominik Kiese}
\affiliation{Center for Computational Quantum Physics, Flatiron Institute, 162 5th Avenue, New York, NY 10010, USA}
\author{Simon Trebst}
\affiliation{Institute for Theoretical Physics, University of Cologne, 50937 Cologne, Germany}
\author{Michael M. Scherer}
\affiliation{Institute for Theoretical Physics III, Ruhr-University Bochum, D-44801 Bochum, Germany}

\begin{abstract}
The discovery of correlated insulating states in moir\'e heterostructures has renewed the interest in strongly-coupled electron systems where spin and valley (or layer) degrees of freedom are intertwined.
In the strong-coupling limit, such systems can be effectively described by SU(4)~spin-valley models akin to Kugel-Khomskii models long studied in the context of spin-orbit coupled materials.
However, typical moir\'e heterostructures also exhibit interactions that break the SU(4)~symmetry down to \svsym.
Here we investigate the impact of such  symmetry-breaking couplings on the magnetic phase diagram for triangular superlattices considering a filling of two electrons (or holes) per moir\'e unit cell.
We explore a broad regime of couplings -- including XXZ anisotropies, Dzyaloshinskii-Moriya exchange and on-site Hund's couplings -- using semi-classical Monte Carlo simulations. 
We find a multitude of classically ordered phases, including (anti-)ferromagnetic, incommensurate, and stripe order, manifesting in different sectors of the spin-valley model's parameter space. 
Zooming in on the regimes where quantum fluctuations are likely to have an effect, we employ pseudo-fermion functional renormalization group (pf-FRG) calculations to resolve quantum disordered ground states such as spin-valley liquids, which we indeed find for certain parameter regimes. 
As a concrete example, we discuss the case of trilayer graphene aligned with hexagonal boron nitride using material-specific parameters.
\end{abstract}\maketitle


\section{Introduction}

Moir\'e systems of different van der Waals heterostructures have by now been established as a highly tunable platform to emulate the physics of strongly correlated electrons~\cite{Cao_2018,2018Natur.556...43C}.
Varying several experimental tuning knobs such as the twist angle between layers, electrical displacement fields and doping via gate voltages, rich phase diagrams have been found to emerge. 
Examples include graphene-based materials such as twisted bilayer graphene (TBG)~\cite{Cao_2018,2018Natur.556...43C,Lu_2019,PhysRevLett.124.076801,Yankowitz_2019,Zondiner_2020,Wong_2020,wolf2021}, twisted double-bilayer graphene (TDBG)~\cite{PhysRevLett.123.197702,Shen_2020,Liu_2020,Cao_2020}, and trilayer graphene aligned with hexagonal boron nitride (TG/h-BN)~\cite{2019NatPh..15..237C,Chen_2019,Chen_2020,Zhou_2021,yang2022}, 
where the occurrence of correlated insulating phases at different integer fillings has been supported by several measurements, sometimes in close proximity with superconducting behavior.
Beyond graphene-based materials, moir\'e bilayers of two-dimensional transition metal dichalcogenides (TMD) have also been found to exhibit correlated states, including Mott insulators, generalized Wigner crystallization, stripe phases, and quantum anomalous Hall insulator phases~\cite{Jin_2019,Regan_2020,Wang_2020,Jin_2021,Li_2021}.

In addition to spin, many of these moir\'e heterostructures feature another bivalued quantum number, which -- depending on the particular system -- has different microscopic origin, e.g., an additional valley or layer index.
Effective descriptions of such heterostructures can then be formulated in terms of Hubbard-type models on a hexagonal moir\'e superlattice, where the electrons come in four flavours due to the combination of the actual spin with the valley/layer pseudospin degree of freedom.
Concretely, models with an approximate SU(4)~flavour symmetry or at least dominating SU(4)-symmetric interactions have been suggested in the context of the above mentioned TBG~\cite{PhysRevLett.121.087001,PhysRevB.98.045103,PhysRevLett.122.246401,PhysRevX.8.031089,Chichinadze_2022}, TDBG~\cite{PhysRevB.99.235406,PhysRevB.102.064501}, TG/h-BN~\cite{zhang2019,PhysRevB.100.035413,PhysRevB.101.035122}, or moir\'e TMDs~\cite{PhysRevLett.127.247701,Xu_2022,PhysRevLett.122.086402,zhang2023}.

The formation of strongly-correlated states observed in various moir\'e systems is tentatively supported by the presence of narrow electron bands, boosting the relevance of electronic interactions.
While it is difficult to pin down the precise location of a moir\'e material on the weak- to  strong-coupling axis, it has been argued that -- building on the four-flavoured Hubbard-type models -- a strong-coupling perspective may be a good starting point to shed light on the nature of the correlated insulated phases at integer fillings, e.g., in TBG~\cite{PhysRevLett.121.087001,PhysRevB.98.045103,PhysRevLett.122.246401}, in  TG/h-BN~\cite{zhang2019,PhysRevB.100.035413,PhysRevX.8.031089} or in moir\'e TMDs~\cite{PhysRevB.100.035413,PhysRevLett.127.247701,zhang2023}.
Such a strong-coupling expansion leads to approximate SU(4) Kugel-Khomskii-type models \cite{kugel1982} for the spin- and the valley/layer degrees of freedom~\cite{PhysRevLett.81.5406,PhysRevLett.81.3527}.
In the context of moir\'e materials, hexagonal-lattice SU(4)~spin models have been previously explored by various many-body approaches, collecting evidence for the emergence of several magnetically ordered as well as spin-valley liquid states~\cite{PhysRevB.100.205131,PhysRevB.100.024421,PhysRevLett.125.117202,keselman2020,kiese2020}.

Accurate modeling of moir\'e materials -- in contrast to paradigmatic studies -- is generally quite challenging.
Generically, however, the approximate SU(4)~symmetry will be broken down to a lower symmetry due to the presence of various competing interaction contributions, see, e.g., Refs.~\cite{PhysRevLett.121.087001,zhang2019,chen2020,PhysRevX.8.031089,PhysRevB.99.195120}, relevant to the cases of TDBG and TG/h-BN.
These interactions can be expected to have a severe impact on the phase diagram by supporting the formation of different spin- and/or valley-ordered magnetic states.
An interesting question therefore is, which of these states are actually realized for specific configurations of SU(4)~breaking couplings.

As a concrete example consider, TG/h-BN with an applied perpendicular electric field $D$, cf.~Ref.~\cite{zhang2019}. 
For large enough $D$ (with a specific sign), topologically trivial, isolated narrow bands emerge that can be described by a spin-valley extended $\mathfrak{su}(4)$\footnote{With $\mathfrak{su}$(4) we refer to the Lie algebra of the Lie group SU(4).} Hubbard model with small anisotropies breaking the SU(4) symmetry. 
This model is expected to undergo a Mott transition when going from smaller to larger fields and may be used to describe the correlated insulating states with $n\!=\!-1$ and $n\!=\!-2$ holes per moir\'e unit cell~\cite{zhang2019,calderon2022,patri2022strong}. Both the type of insulator in the strong coupling limit, as well as the nature of the Mott transition, are still under debate. 
In Ref.~\cite{zhang2019}, where a concrete spin-valley extended Hubbard model for TG/h-BN was derived, the insulating phase was predicted to be ferromagnetic in the strong-coupling limit (Equation \eqref{eq:hamiltonian} below), with a possible spin liquid phase close to the Mott transition. 
More recent work considered a Hubbard model with a completely SU(4)~symmetric onsite interaction using single-site DMFT~\cite{calderon2022}. 
Their results suggest that close to the Mott transition interactions promote antiferromagnetic order, which breaks $C_3$ symmetry and competes with the ferromagnetic state. 
This seems to be compatible with spectroscopy measurements of TG/h-BN observing a direct optical excitation across the Mott gap that rules out a ferromagnetic ground state, but allows antiferromagnetic or intervalley-coherent ground-state order~\cite{yang2022}.

For the case of TG/h-BN and also for other moir\'e heterostructures with intertwined spin and valley/layer degrees of freedom, the additional SU(4)~breaking couplings may, however, not be negligible, and could, thus, strongly affect the nature of correlated states of the system.
We therefore deem it useful to gain a more general understanding of effective models in different parameter regimes and limits.
Here, we shed some light on this issue, focusing on moir\'e systems that can be described on a triangular superlattice, with a filling of two electrons (or holes) per moir\'e unit cell, as relevant, e.g., for TDBG, TG/h-BN, and moir\'e TMDs.
To this end, we study the limit of strong coupling, i.e. we explore a spin-valley model with localized degrees of freedom described by generators of SU(4), featuring both an on-site Hund's-type coupling as well as nearest and next-nearest neighbor superexchange interactions that strongly break the SU(4) symmetry down to \svsym.
Due to the extra valley degeneracy, the size of the local Hilbert space in spin-valley extended models is significantly increased compared to conventional Mott insulators, which makes the theoretical treatment a serious challenge for most conventional many-body techniques.

We start to explore such models by using a semi-classical Monte Carlo approach to establish the classical ground-state order in the high-dimensional SU(4) spin-valley space, similar to what was already done analytically for the fully SU(4)~symmetric model~\cite{keselman2020}. Since the parameter space in low-symmetric models is large, we here take a concretely suggested model for TG/h-BN as reference and explore the parameter space and phase diagram in its vicinity to classify candidates for the insulating phases of TG/h-BN and, possibly, related moir\'e systems.  In doing so, we find a multitude of classically ordered phases, including antiferromagnetic (AFM), ferromagnetic (FM), incommensurate (ICS) and stripe order that manifest in different sectors of the coupled spin-valley space. It turns out that a determining factor in the ground-state order are indeed the type and strength of the SU(4) symmetry breaking interactions, which we systematically investigate. We further study the effect of thermal fluctuations and find evidence for both continuous and first-order transitions into the ordered phases, as well as order-by-disorder effects where thermal fluctuations lift the ground-state manifold.

To also study the effect of quantum fluctuations, which are expected to be especially strong in the vicinity of the Mott transition, we complement our semi-classical analysis by performing pseudo-fermion functional renormalization group (pf-FRG) calculations~\cite{woelfle2010} to distinguish between magnetically ordered and disordered regimes.
The pf-FRG approach has recently made considerable progress in broadening its range of describable models~\cite{gresista2022}.
Concretely, first applications of the pf-FRG to SU(N)~symmetric spin models~\cite{PhysRevB.97.064415,PhysRevB.97.064416} have been extended to the case of SU(2)${}_{\mathrm{spin}}\otimes$SU(2)${}_{\mathrm{valley}}$ symmetry~\cite{kiese2020} and further technical developments have prepared applicability to general \svsym\ models~\cite{gresista2022}. 
Here, we will apply an implementation of the latter methodological refinement to the specific case of spin-valley models for moir\'e heterostructures, which gives us estimates for the location of putative spin-valley liquid or other quantum disordered phases in parameter space.

The rest of the manuscript is structured as follows: 
We start with short overview over the key results of our numerical study in Sec.~\ref{sec:key-results}, focusing on phase diagram for the strong-coupling limit of the spin-valley model derived in Ref.~\cite{zhang2019} for TG/h-BN. In Sec.~\ref{sec:model}, we introduce a general class of spin-valley models with \svsym\ symmetry and also briefly comment on their possible origin in the context of moir\'e heterostructures. 
In Sec.~\ref{sec:semiclass}, we then introduce the semi-classical Monte Carlo approach to study the spin-valley model and analyze its ordered phases with a focus on a region in parameter space suggested for TG/h-BN and beyond. 
We specifically also address the role of SU(4)-symmetry breaking couplings within the semi-classical analysis.
To study the role of quantum fluctuations, in Sec.~\ref{sec:frg}, we employ an extension of the pf-FRG approach to identify regions in the phase diagram where quantum disordered states, spin-valley liquids, can occur.
We conclude with a discussion in Sec.~\ref{sec:discussion}.


\section{Key results}\label{sec:key-results}

\begin{figure}[t!]
    \centering
    \includegraphics{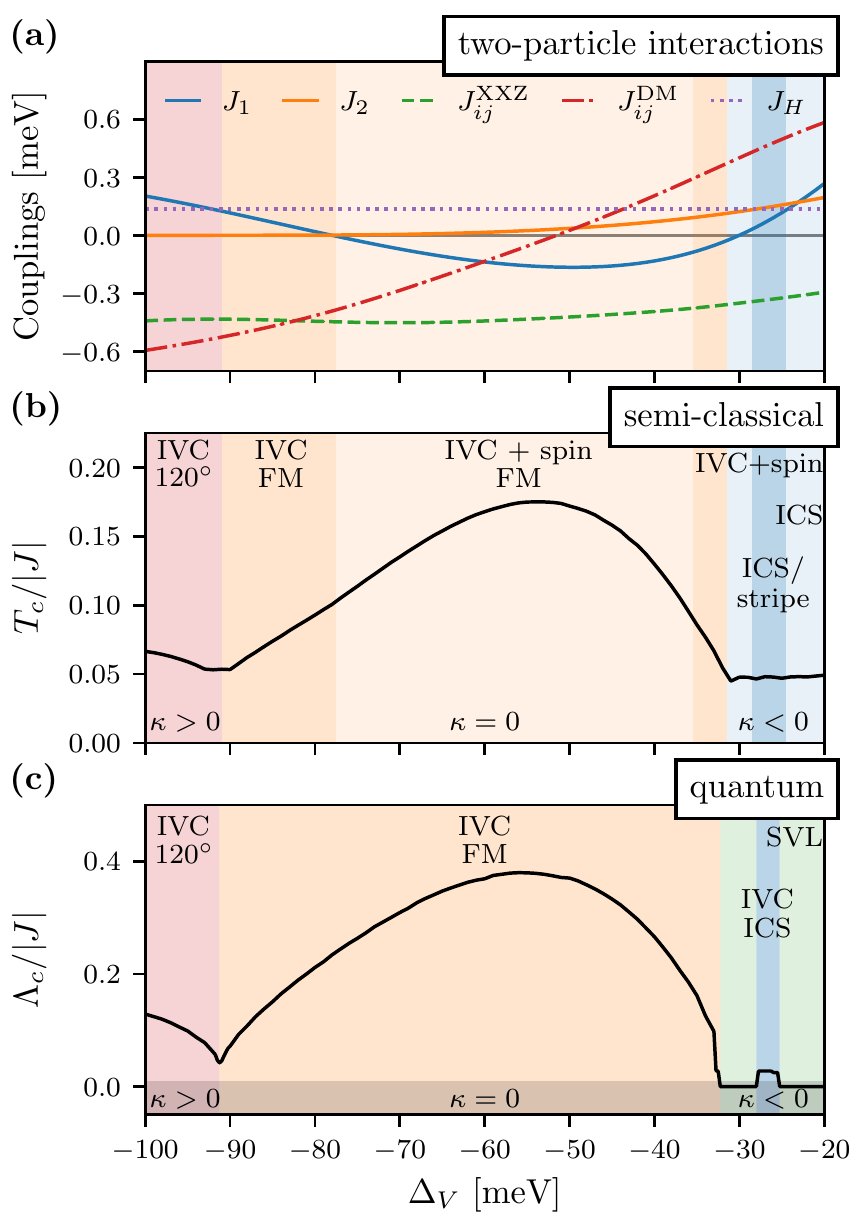}
    \caption{{\bf Phase diagram of the TG/h-BN model} as a function of the potential difference $\Delta_V$ with
    \textbf{(a)} the couplings in Hamiltonian~\eqref{eq:hamiltonian} as estimated for TG/h-BN \cite{zhang2019}, \textbf{(b)} the transition temperature $T_c$ and ground-state order from semi-classical Monte Carlo calculations and \textbf{(c)} the critical scale and quantum ground states from pf-FRG. Away from the Mott transition at $\Delta_V = -20\mev$ the ground state is ferromagnetic (FM) or $120^\circ$ ordered in both the semi-classical and quantum limit. Close to the Mott transition, the semi-classical calculation shows incommensurate (ICS) order that melts into a quantum disordered, putative spin-valley liquid (SVL) state when including quantum fluctuations, as indicated by the absence of a flow breakdown in the {pf-FRG}  ($\Lambda_c = 0$). In the semi-classical phase labeled by ICS/stripe, the valley shows ICS order and the spin instead shows stripe order.  The pf-FRG has difficulties to resolve phases with order in multiple sectors and only predicts the dominant IVC order. The $120^\circ$ and ICS ordered states have a finite vector chirality $\kappa$ -- defined in Eqs.~\eqref{eq:chirality} and~\eqref{eq:specific-chiralities} -- that switches sign in line with the sign change in $J_{ij}^{\mathrm{DM}}$. All observed ground states break the $\mathrm{U(1)}_\mathrm{valley}$ symmetry implying inter-valley coherent (IVC) order, and, in some cases, additionally the spin shows finite expectation values (IVC + spin).}
    \label{fig:tghbn-phase-diagram}
\end{figure}

We begin by giving a short overview over the main results from our numerical study of the \svsym \ symmetric spin-valley model (Eq.~\eqref{eq:hamiltonian} below) originally derived in Ref.~\cite{zhang2019} for TG/h-BN in the strong coupling limit. We always consider a filling of $n=-2$ holes per moir\'e unit cell. A more detailed introduction to the model, our numerical methods, and an in-depth discussion of the results are presented in the following sections.

Starting with the concrete example of TG/h-BN, the applied displacement field $D$ induces a potential difference $\Delta_V$ between the graphene layers, which tunes both the overall strength of electronic correlations, as well as the relative strength between different types of interactions. Ref.~\cite{zhang2019} predicts that the strongest electronic correlations  occur for $\Delta_V < - 30\mev$, while for $\Delta_V > - 30\mev$ their strength decreases until at approximately $\Delta_V \simeq -20\mev$ a Mott transition to a metallic state occurs. The dependence of the couplings in the strong-coupling Hamiltonian on the potential difference is shown in Fig.~\ref{fig:tghbn-phase-diagram}~(a), according to Ref.~\cite{zhang2019}. The Hamiltonian includes SU(4) symmetric nearest and next-nearest neighbor interactions $J_1$ and $J_2$, as well as SU(4) breaking interactions in the form of an XXZ anisotropy $J^\mathrm{XXZ}_{ij}$, a Dzyaloshinskii-Moriya exchange $J_{ij}^\mathrm{DM}$ and an on-site Hund's couplings $J_H$. 

Fig.~\ref{fig:tghbn-phase-diagram}~(b) and (c) show the corresponding semi-classical and quantum phase diagrams obtained from our Monte Carlo and pf-FRG calculations, respectively, as presented in this work. Away from the Mott transition the ground state is an ordered state exhibiting either ferromagnetic or $120^\circ$ order in both the semi-classical and quantum limit. Close to the Mott transition, however, the semi-classical approach shows incommensurate (ICS) order that melts into a  quantum disordered, putative spin-valley liquid (SVL) state when including quantum fluctuations. All observed ground states break the $\mathrm{U(1)}_\mathrm{valley}$ symmetry implying inter-valley coherent (IVC) order. We never observe valley polarization, i.e. finite $\tau^z$. In some cases, the spin shows finite expectation values in addition to IVC order, which we refer to as `IVC + spin' order. In the quantum case, however, the pf-FRG has difficulties to resolve such spin-valley order in multiple sectors and only shows the more dominant IVC order.

Fig.~\ref{fig:tghbn-phase-diagram} suggests that the type of ground-state order strongly depends on the value of the SU(4) breaking couplings. A dominant $J^\mathrm{XXZ}_{ij}$ results in collinear, ferromagnetic order, while large values of $J_{ij}^\mathrm{DM}$ seem to induce non-collinear order with a finite vector chirality $\kappa$ -- defined in Eqs.~\eqref{eq:chirality} and~\eqref{eq:specific-chiralities} -- such as $120^\circ$ or ICS order. The type of non-collinear order then depends on the relative magnitude of $J_1$ and $J_2$. A small $J_2/J_1$ favors $120^\circ$ order and a large $J_2/J_1$ favors ICS order. We confirm this observation by systematically varying the SU(4) breaking couplings for different (fixed) values of $J_1$ and $J_2$, as is shown in Fig.~\ref{fig:quantum_phase_diagram_flux}. The quantum limit agrees with the semi-classical case in all but the ICS phases, where we consistently observe that quantum fluctuations drive the system into a disordered state.

Our two complementary approaches, targeting both classically ordered and quantum disordered phases, provide a consistent picture of the spin-valley entangled quantum magnetism across a wide range of coupling parameters. 
While our analysis, originating from a strong coupling approach, is per se agnostic to the details of the electronic state, one might nevertheless be tempted to make some connections. 
The strong-coupling perspective is best justified for correlated insulator states, particularly Mott insulating states with local spin and valley degrees of freedom (which is the case for large displacement fields $\Delta_V < -30$~meV). 
In this regime, the classically ordered states that we find -- various forms of ferromagnetism and 120$^\circ$ order -- are good candidates for the collective spin-valley ordered ground state.
In the weakly correlated regime for $\Delta_V \gtrsim -30$~meV, our strong-coupling approach is less justified. 
But it is remarkable that it is in this regime where we find the emergence of quantum disordered ground states, even in the presence of sizable SU(4)~breaking couplings. 
The formation of such spin-valley quantum liquids in the weakly coupled regime is reminiscent of the observation that quantum spin liquids~\cite{Balents2010,Savary2016} might form in weak Mott insulators close to the metal-insulator transition (where electronic fluctuations might imprint themselves onto the magnetic state).
Our results might therefore motivate further searches for spin-liquid states in strongly-correlated moir\'e heterostructures.


\section{Spin-valley model}\label{sec:model}

Starting from a spin-valley extended Hubbard model, a strong coupling expansion leads to a Kugel-Khomskii type~\cite{kugel1982} spin-valley model. 
Therein, instead of conventional $\mathfrak{su}(2)$ spin operators, the 15~generators of SU(4), forming the basis of the Lie algebra $\mathfrak{su}(4)$, describe the localized degrees of freedom.
In the context of moir\'e materials it can be useful to keep the original spin and valley quantum numbers explicit and choose a basis of \emph{spin-valley operators} defined via a parton construction with auxiliary pseudo-fermions as\footnote{Summation over repeated spin and valley indices is implied.}
\begin{align}
    \sigma_{i}^{\mu} \tau_{i}^{\kappa} 
    &= f_{i s l}^{\dagger} \theta_{s s'}^{\mu} \theta^{\kappa}_{l l'} f^{\pdagger}_{i s' l'}\,, \notag \\
    \sigma_i^{\mu} \equiv \sigma_i^{\mu} \mathds{1}_i 
    &= f_{i s l}^{\dagger} \theta_{s s'}^{\mu} f^{\pdagger}_{i s' l}\,, \notag \\
    \tau_i^\kappa \equiv \mathds{1}_i\tau_i^{\kappa} 
    &= f_{i s l}^{\dagger} \theta_{l l'}^{\kappa} f^{\pdagger}_{i s l'} \,,
    \label{eq:spin-valley-operators}
\end{align}
where  $f_{i s l}^{\dagger}$ and $f^{\pdagger}_{i s l}$ are fermionic creation and annihilation operators with a site index $i$, a spin index $s = (\uparrow, \downarrow)$, and a valley/layer index $l = (+, -)$.
The $\theta^\mu$ $(\mu = x, y, z)$  are the usual Pauli matrices. 
In the following, we will simply call $l$ the ``valley index" and note that it is determined by the specific system under consideration, whether it refers to an actual valley, a layer, or another pseudospin degree of freedom.

We want to consider the scenario of half electron (or hole) filling of the underlying Hubbard model, which translates to a local constraint of two partons per site, i.e.
\begin{equation}
  n_i = f_{i sl}^{\dagger} f_{i sl}^{\pdagger} = 2\,.
  \label{eq:half-filling}
\end{equation}
With four electron flavours given by the different combinations of the spin and valley indices, this corresponds to half-filling and fixes the local Hilbert space dimension to $\binom{4}{2} = 6$, which means we consider the six-dimensional representation of $\mathfrak{su}(4)$. 
The corresponding SU(4) symmetric Heisenberg model with antiferromagnetic couplings on a triangular lattice has been studied previously, exhibiting a non-magnetic valence bond solid (VBS) ground state~\cite{keselman2020}.

For many moir\'e materials, SU(4) breaking terms due to on-site and inter-site Hund's coupling are estimated to be small compared to density-density interactions and using an SU(4)~symmetric interaction can be justified \cite{zhang2019, calderon2022, chen2020}. 
In the strong coupling limit, however, the perturbative treatment of the Hamiltonian's kinetic term generates superexchange interactions that strongly break the SU(4) symmetry and are, in fact, of the same order as the SU(4) symmetric terms \cite{venderbos2018, zhang2019}. 
Here, we therefore consider a spin-valley model defined by the Hamiltonian
\begin{equation}
\label{eq:hamiltonian}
    \begin{aligned}
        H &= \frac{J_1}{8} \sum_{\langle ij \rangle} (1 + \boldsymbol{\sigma}_i \boldsymbol{\sigma}_j)(1 + \boldsymbol{\tau}_i \boldsymbol{\tau}_j) \\[5pt]
        &+ \frac{J_2}{8} \sum_{\langle\!\langle ij \rangle\!\rangle} (1 + \boldsymbol{\sigma}_i \boldsymbol{\sigma}_j)(1 + \boldsymbol{\tau}_i \boldsymbol{\tau}_j) \\[5pt]
        &+ \frac{1}{8} \sum_{\langle ij \rangle} J_{ij}^\mathrm{XXZ}(1 + \boldsymbol{\sigma}_i \boldsymbol{\sigma}_j)(\tau^x_i \tau^x_j +\tau^y_i \tau^y_j) \\[5pt]
        &+ \frac{1}{8} \sum_{\langle ij \rangle} J^\mathrm{DM}_{ij}(1 + \boldsymbol{\sigma}_i \boldsymbol{\sigma}_j)(\tau^x_i \tau^y_j - \tau^y_i \tau^x_j) \\[5pt]
        &- \frac{J_H}{4} \sum_i \left(n_{+i}n_{-i} + \boldsymbol{\sigma}_{+i}\boldsymbol{\sigma}_{-i}\right)\,,
    \end{aligned}
\end{equation}
where $\langle\cdot\rangle$ denotes summation over nearest-neighbor sites and $\langle\!\langle\cdot\rangle\!\rangle$ over next-nearest neighbors, in our case of a triangular lattice. 
Such a Hamiltonian can be obtained starting from an extended spin-valley Hubbard model for TG/h-BN \cite{zhang2019} and then using standard second-order perturbation theory. 

The terms proportional to $J_1$ and $J_2$ are SU(4) symmetric nearest and next-nearest neighbor Heisenberg interactions. 
The coupling $J_1$ can include both a ferromagnetic part arising from a nearest-neighbor Hund's coupling and an antiferromagnetic part due to superexchange. 
Depending on the precise material parameters and the strength of the applied displacement field, both positive and negative couplings are thus possible, see, e.g., the discussion of the specific case of TG/h-BN, below. 
The coupling $J_2$, in contrast,  can be expected to be positive as it exclusively originates from a superexchange process. 
The coupling terms proportional to $J_{ij}^\mathrm{XXZ}$ and $J_{ij}^\mathrm{DM}$ break the SU(4) symmetry down to \svsym\ and also mainly originate from superexchange. 

In addition to nearest and next-nearest neighbor interactions, the last term in the Hamiltonian is an on-site Hund's coupling $J_H$, where $n_{\pm i} = f_{i\pm l}^{\dagger} f_{i\pm l}^{\pdagger}$ and {$\sigma^\mu_{\pm i} = f_{i s \pm }^{\dagger} \theta_{s s'}^{\mu} f^{\pdagger}_{i s' \pm }$} are the density and spin operators in the $\pm$ valley sectors. 
This term can be rewritten, up to a constant, in terms of the spin-valley operators in Eq.~\eqref{eq:spin-valley-operators} using
\begin{equation}
    \begin{aligned}
         n_{+i}n_{-i} + \boldsymbol{\sigma}_{+i}\boldsymbol{\sigma}_{-i} =  \frac{1}{4}(1 + \boldsymbol{\sigma}_i \boldsymbol{\sigma}_i)(1-\tau_i^z \tau_i^z)\,,
    \end{aligned}
\end{equation}
which shows more clearly that this also breaks the SU(4) symmetry to \svsym.

In total, this model introduces five independent coupling parameters, i.e. 
\begin{align}
J_1, J_2, J_{ij}^\mathrm{XXZ}, J_{ij}^\mathrm{DM}\ \text{and}\ J_H\,,
\end{align}
that may all be of similar relevance. 
According to the motivation given, specific choices for these parameters describe the strong-coupling sector of various moir\'e heterostructures.
\subsection{TG/h-BN spin-valley model}

Since such a huge parameter space is hard to explore exhaustively, we use the estimated values for TG/h-BN from Zhang and Senthil's model building in Ref.~\cite{zhang2019} as a starting point and then explore the parameter space in its vicinity.
For TG/h-BN, $J_{ij}^\mathrm{XXZ}$ and $J_{ij}^\mathrm{DM}$ are related to  $J_1$ via 
\begin{equation}
\begin{aligned}
    J_{ij}^\mathrm{XXZ} &= (J_1 + K)[\cos{(2\varphi_{ij})} - 1]\,,\\
    J^\mathrm{DM}_{ij} &= (J_1 + K)\sin{(2\varphi_{ij})} \,.
\end{aligned}
\label{eq:su4-breaking-couplings}
\end{equation}
Here, $\varphi_{ij}$ is a phase that flips sign between nearest-neighbor bonds that are related by a $C_6$ rotation, hence breaking the $C_6$ symmetry of the triangular lattice down to $C_3$. 
This phase originates from a valley contrasting flux  $|\Phi| = 3|\varphi_{ij}|$ in the nearest-neighbor hopping. 
$K$ is an intersite Hund's coupling. 

Notably, for TG/h-BN, the parameters $J_1, J_2$ and $\varphi \equiv |\varphi_{ij}|$ are {\it tunable} by an applied displacement field, which induces a potential difference $\Delta_V$ between the graphene layers. 
The dependence on this potential difference is depicted in Fig.~\ref{fig:tghbn-phase-diagram}(a). 
$K$ and $J_H$ only depend weakly on the displacement field and we consider the constant values $K = 0.4\mev$ and $J_H = 0.136\mev$ in the following.
Note that Ref.~\cite{zhang2019} predicts that the size of the Hubbard $U$ compared to the electronic bandwidth is strongest for $\Delta_V < - 30\mev$, i.e. this regime is the most likely to be appropriately described by a strong-coupling expansion.
In contrast, $U$ becomes equal to the bandwidth at $\Delta_V \simeq -20\mev$, where the Mott transition is expected to occur. 
Approaching such a Mott transition from the insulating side, it is expected that the spin-valley model at hand will be further augmented by higher-order exchange terms, which become relevant in the strong coupling expansion.
Here, we adopt a perspective that the spin-valley model~\eqref{eq:hamiltonian} is a well-justified starting point for the case of TG/h-BN, 
cf. Refs.~\cite{zhang2019,patri2022strong}, but might also be an interesting model in its own right. We will explore the full range 
of coupling parameters, including the vicinity of the Mott transition where new physics might arise.


\section{Semi-classical analysis}\label{sec:semiclass}

To understand the role of the different couplings in our principal Hamiltonian~\eqref{eq:hamiltonian} it is useful to start with a classical analysis and study which kind of ordered ground states are favored by the different interactions in the classical limit. 
In this section, we do so by first  defining a suitable (semi)-classical limit for the $\mathfrak{su}(4)$ models in consideration and then discuss how we study the resulting model using classical Monte Carlo calculations.
This allows us not only to determine the ground-state order over vast parameter ranges, but also to study the role of thermal fluctuations and the nature of thermal phase transitions into these ordered phases.

\subsection{Semi-classical Monte Carlo}
For $\mathfrak{su}(2)$ spin models the usual classical limit is to replace the spin operator by real three-dimensional vectors of fixed length, which has proven useful for a multitude of quantum spin models even for small spin lengths $S$ \cite{LandauBinder}. 
A naive continuation of this method to spin-valley models would be to perform a mean-field decoupling of the spin and valley degree of freedom into two separate $\mathfrak{su}(2)$ spins, and then approximate these spins by classical vectors.
Although such an approach might give reasonable results for a filling of one electron per site \cite{venderbos2018, xian2021}, at two electrons per site spin- and valley degrees of freedom are closely intertwined, as e.g. $\langle \tau_i^z \rangle = 2$ directly implies $\langle \sigma^z_i \rangle = 0$ due to the Pauli principle. 

Such a naive mean-field decoupling of spin and valley is, therefore, not an ideal approximation beyond single-electron filling. 
Instead, we follow Ref.~\cite{keselman2020} and define the semi-classical limit solely by the fact that there is no entanglement between two different lattice sites. 
This is enforced by considering only product states of the form 
$|\psi\rangle = \otimes_{i} |\psi_i\rangle$, where $|\psi_i\rangle$ is an arbitrary state in the local Hilbert space on site $i$. In order to find the semi-classical ground state at $T = 0$, we minimize the semi-classical energy $H_\mathrm{sc} = \langle\psi| H |\psi\rangle$ numerically using a simulated annealing procedure with subsequent stochastic gradient descent (details of which are provided in appendix~\ref{app:mc}). 
On a conceptual level, this approach is equivalent to a mean-field theory where spin-valley operators $\sigma_i\tau_i$ are replaced by their expectation values $\langle \sigma_i\tau_i\rangle$ and determined self-consistently. 

We can, however, go one step further and also calculate expectation vales at finite temperatures $T = 1/\beta >0$ by sampling the space of product-state wavefunctions according to the Boltzmann distribution $\sim \exp(-\beta \langle\psi| H |\psi\rangle)$ ~\cite{stoudenmire2009, hickey2014} using a standard Markov chain Monte Carlo algorithm \cite{LandauBinder}. This is equivalent to a cumulant expansion of the partition function to first order, which becomes exact in the limits of low $T \to 0$ and high temperature $T \to \infty$ and we therefore expect it to accurately capture the thermodynamic properties of the semi-classical model. Further details on the simulations are also given in appendix~\ref{app:mc}.

\subsection{Ordered phases of the spin-valley model}

As a starting point for our analysis, we calculate the ground-state and finite-temperature phase diagram for the Hamiltonian~\eqref{eq:hamiltonian} with the coupling parameters given by the estimates for TG/h-BN in Ref.~\cite{zhang2019}. 
The dependence of these couplings parameters as a function of the potential difference $\Delta_V$ is shown Fig.~\ref{fig:tghbn-phase-diagram}~(a), while Fig.~\ref{fig:tghbn-phase-diagram}~(b) shows a summary of the finite-temperature phase diagram {(using $|J| = (J_1^2 + J_2^2 + {J_{ij}^\mathrm{XXZ}}^2 + {J_{ij}^\mathrm{DM}}^2 + J_H^2)^{1/2}$ as a an energy scale for normalization)}.

\begin{figure}
    \centering 
    \includegraphics{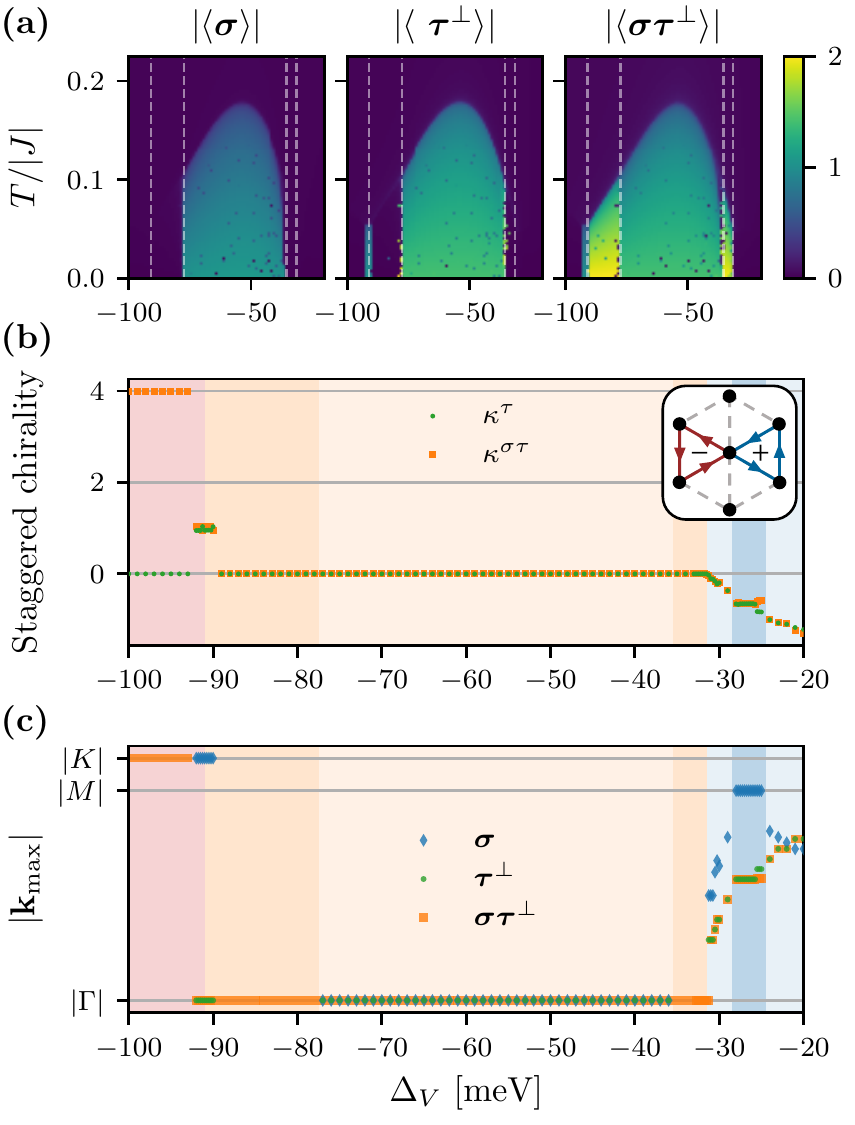}
    \caption{{\bf Semi-classical observables} as a function of the potential difference $\Delta_V$. \textbf{(a)} Magnetization in the spin $\bsigma$, the in-plane valley $\btaup$ and the in-plane spin-valley $\bsigma\btaup$. Finite magnetization implies FM order in the respective sector. A crossover between IVC FM order (only finite $|\langle\bsigma\btau\rangle|$) and IVC + spin FM order (all magnetizations finite) can be observed. The dashed lines show the phase boundaries. \textbf{(b)}~Staggered vector chirality of the ground state as defined in Eqs.~(\ref{eq:chirality}, \ref{eq:specific-chiralities}) and illustrated in the inset. At low $\Delta_V$ the value  $\kappa^{\sigma\tau} = +4$ implies IVC $120^\circ$ order. The varying negative $\kappa^{\tau}\approx\kappa^{\sigma\tau}$ at higher $\Delta_V$ implies IVC + spin ICS or stripe order. \textbf{(c)} Magnitude of the ground-state ordering vector corresponding to the position of the structure factor's maximum $\mathbf{k}_\mathrm{max}$. At each $\Delta_V$ only sectors that show significant non-zero correlations are shown. $\mathbf{k}_\mathrm{max} = K (K')$ corresponds to $120^\circ$ order, $\mathbf{k}_\mathrm{max} = \Gamma$ to FM order, $\mathbf{k}_\mathrm{max} = M$ to stripe order and values on no symmetry point of the Brillouin zone to ICS order. Examples of the full structure factors are shown in Fig.~\ref{fig:classical_sfs_field}.}
    \label{fig:classical_observables}
\end{figure}

\begin{figure*}
    \centering
    \includegraphics{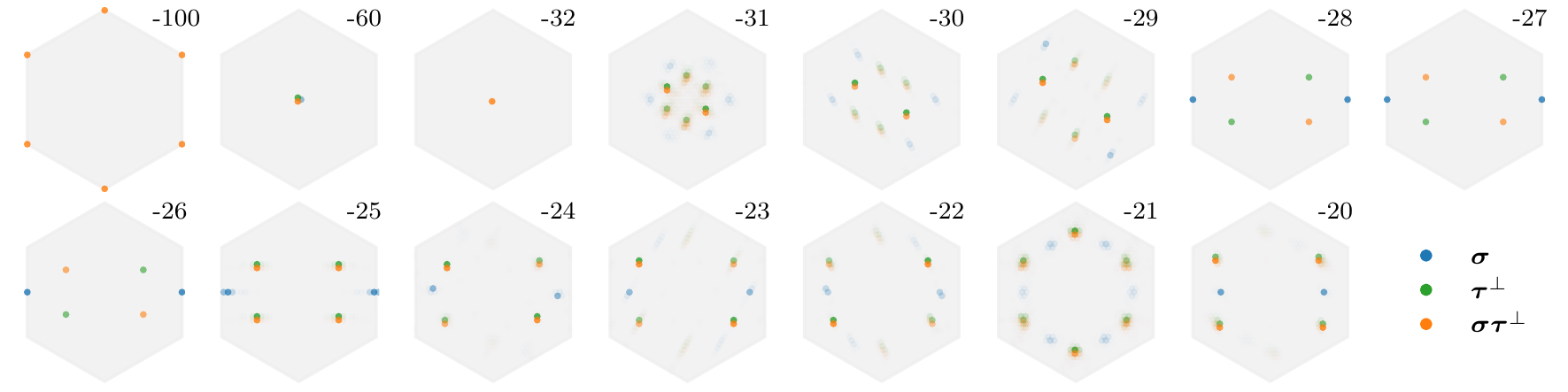}
    \caption{{\bf Semi-classical ground-state structure factors} (static) for different $\Delta_V$ (in meV). The color and opacity depict the sector and the relative magnitude, respectively. Points that lie on top of each other are slightly shifted for visibility.}
    \label{fig:classical_sfs_field}
\end{figure*}

\subsubsection{Preliminaries}

Before we discuss the different phases for specific parameter choices, let us briefly review the types of order one can generally expect for the model at hand. 
The simplest order is of one-sublattice type, i.e.\ ferromagnetic. 
For $\mathfrak{su}(2)$ spin models, this means that all spins are aligned and point somewhere on the 2-sphere in spin space. 
The analogue for $\mathfrak{su}(4)$ models is a state for which the expectation values of the 15 generators $\sigma^\mu, \tau^\nu, \sigma^\mu\tau^\nu$ are the same on all sites. 
Since the Hamiltonian we consider is symmetric under global rotations, generated by $\boldsymbol{\sigma}$, $\tau^z$ and $\boldsymbol{\sigma} \tau^z$, the states related by these rotations have the same energy. 
Therefore, we do not distinguish between order in all 15 generators, but we define the following sectors
\begin{equation}
    \begin{aligned}
        \bsigma &= (\sigma^x, \sigma^y, \sigma^z) \,, \\
        \btaup &= (\tau^x, \tau^y) \,,\\
        \tau^z &= (\tau^z)\,, \\
        \bsigma\btaup &= (\sigma^x\tau^x, \sigma^y\tau^x, \sigma^z\tau^x, \sigma^x\tau^y, \sigma^y\tau^y, \sigma^z\tau^y)\,,\\
        \bsigma\tau^z &= (\sigma^x\tau^z, \sigma^y\tau^z, \sigma^z\tau^z)\,.
    \end{aligned}
    \label{eq:sectors}
\end{equation} 
Since all generators have eigenvalues -2, 0 and 2, a perfect ferromagnet that orders, e.g., only in the spin $\bsigma$ will have a finite magnetization $|\langle \bsigma \rangle| \equiv |\sum_i \langle \bsigma_i \rangle|/N= 2$ with the magnetization in all other sectors precisely vanishing\footnote{Note that this is quite different to the case of quarter filling (one electron per site), where different sectors can have maximal magnetization simultaneously, e.g., showing complete spin and valley polarization.}. 

Given that our model exhibits a continuous SU(2)$_\mathrm{spin}$ symmetry, a fluctuation-driven finite-temperature transition into pure spin order  is generally forbidden by the Mermin-Wagner theorem \cite{Mermin-1966}. 
We, therefore, expect order either in the valley or spin-valley sectors. Order in the out-of-plane valley $\tau^z$ or spin-valley $\bsigma\tau^z$ implies a valley polarized state that breaks an Ising like $Z_2$ symmetry. 
This is, however, not observed in any of the calculations in this work. Order in the in-plane valley $\btaup$ or spin-valley  $\bsigma \btaup$ breaks the continuous U(1)$_\mathrm{valley}$ symmetry implying \emph{inter-valley coherent} (IVC) order and may be realized through a Berezinskii–Kosterlitz–Thouless type transition \cite{berezinsky1970, kosterlitz1973}. 
In the following, we label ordered states that only have finite $\langle\btaup\rangle$ or $\langle\bsigma\btaup\rangle$ as `IVC', states that additionally have $\langle\bsigma\rangle\neq 0$ as `IVC + spin' and states that only have finite $\langle\bsigma\rangle$ as `spin'.

In addition to ferromagnetic order, the system may also realize states where the expectation values of the $\mathfrak{su}(4)$ generators vary spatially, e.g. in a $120^\circ$ type or even an incommensurate (ICS) pattern. This variation may take place in only one of the defined sectors or, in principle, in the full $\mathfrak{su}(4)$ spin-valley space. To fully label an ordered state, we therefore have to give both the sector the order occurs in, as well as the type of real-space pattern. As an example, if a state shows a $120^\circ$ pattern in $\langle\bsigma\btaup\rangle$, with all other expectation values being zero, we refer to it as IVC $120^\circ$ order.

\subsubsection{Ferromagnetic states}

For a large region of the phase diagram, centered at around $\Delta_V = -60\mev$, we obtain ferromagnetic order showing finite magnetizations in different sectors, as shown in Fig.~\ref{fig:classical_observables}~(a). 
In this regime, the out-of-plane (spin-)valley expectation values all vanish (not shown),  which is not surprising, given that the couplings $J_1$ and $J_{ij}^\mathrm{XXZ}$ are dominant and negative here, favoring ferromagnetic order in $\btaup$ and $\bsigma\btaup$.

Interestingly, we find {\it two} different kinds of ferromagnets: 
At the boundaries of the FM phase, i.e. in the regions $-91 \lesssim \Delta_V \lesssim -77 \mev$ and $-36 \lesssim \Delta_V \lesssim -31 \mev$, we see IVC FM order and the ground state is simply an eigenstate of $\btaup$ \textit{or} $\bsigma\btaup$. Thermal fluctuations always select an eigenstate of $\bsigma\btaup$ and not $\btaup$ indicating an order-by-disorder effect. 
In the intermediate range $-77 \lesssim \Delta_V \lesssim -36 \mev$ the ordered state is an IVC + spin FM and has finite magnetizations in both $\bsigma$, $\btaup$ and $\bsigma\btaup$. 
Here, up to symmetry transformations, the ground state can be written as 
\begin{equation}
\label{eq:mixed-state}
    |\Psi\rangle^\mathrm{IVC+spin}_\mathrm{FM} \sim |\sigma^x\tau^x\rangle + |\tau^x \rangle + \delta |\sigma^x\rangle \,,
\end{equation}
where by $|\sigma^\mu\tau^\nu\rangle$ we denote the eigenstate of $\sigma^\mu\tau^\nu$ with eigenvalue $+2$. 
The  value of $\delta$ varies as a function of $\Delta_V$ in the interval $\delta \in [0.455, 0.538] \mev$ which we found by comparing the analytic energy of $|\Psi\rangle^\mathrm{IVC+spin}_\mathrm{FM}$  with the numerical minimization. 
We note that since the eigenstates $|\sigma^x\tau^x\rangle$, $|\tau^x \rangle$ and $|\sigma^x\rangle$ are not linearly independent, their linear combination may also be written differently and already the state $|\sigma^x\tau^x\rangle + |\tau^x \rangle$ would generate a finite spin expectation value $\langle\sigma^x\rangle$, which lowers the energy of the on-site Hund's coupling term $\sim J_H$.

\begin{figure*}
    \centering
    \includegraphics{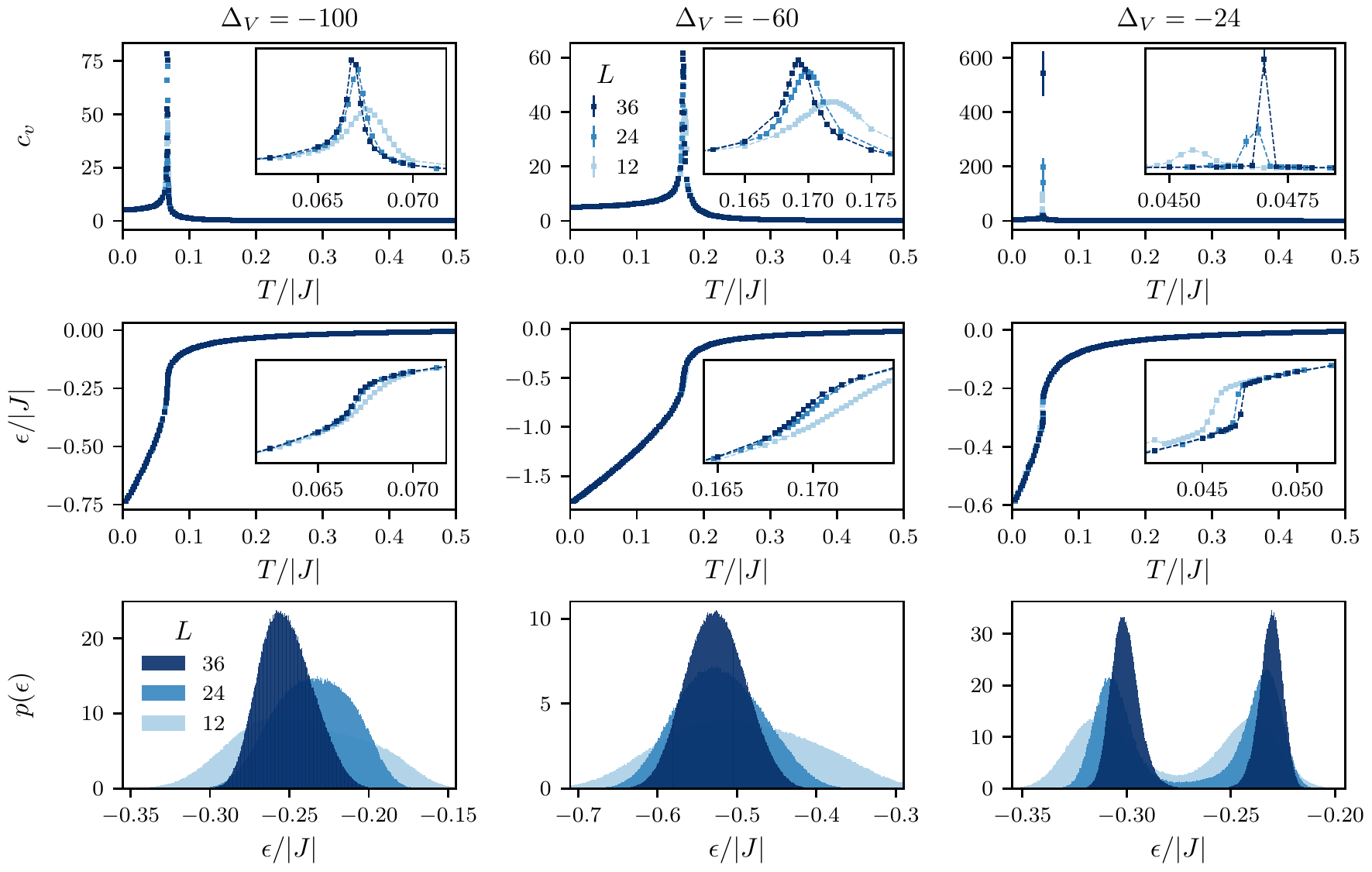}
    \caption{{\bf Thermodynamics of the spin-valley model} in semi-classical Monte Carlo simulations for TG/h-BN inspired coupling parameters
			shown for three different potential differences $\Delta_V = -100, -60, -24$~meV, 
			which stabilize a ($120^\circ$) ordered state (left column), a spin-valley ferromagnet (middle column), 
			and incommensurate (ICS) order (right column).
			The top row shows the specific heat, the middle row the energy per site, and the bottom row an energy histogram
			at the thermal phase transition. The double-peak structure in the latter indicates a first-order transition.}
    \label{fig:thermodynamics-field}
\end{figure*}

\subsubsection{Ordered states with finite chirality}

 Outside the ferromagnetic region all one-sublattice magnetizations vanish and other observables need to be considered to classify the phases. 
 Here the Dzyaloshinskii–Moriya-type coupling $J^{DM}_{ij}$ is large which favors non-collinear states with a finite chirality in $\btaup$ and/or $\bsigma\btaup$. 
 This means that, going around a triangle, expectation values will likely be rotated in valley space relative to their neighbors with a certain handedness that is opposite for left and right pointing triangles. 
 The prime example with maximal chirality is $120^\circ$ order. 
 To quantify this, we calculate the $z$-component of the staggered vector chirality, which for three dimensional vectors $\mathbf{v}_i$ distributed on the triangular lattice is defined as {(see Fig.~\ref{fig:classical_observables}~(b) for an illustration)}
\begin{equation}
    \kappa(\mathbf{v}) = \frac{1}{3\sqrt{3}N}\!\!\sum_{r \in (\triangleright, \triangleleft)}\!\!\!(-1)^r\!
    \left(\mathbf{v}_1\!\times\!\mathbf{v}_2 + \mathbf{v}_2\!\times\!\mathbf{v}_3 + \mathbf{v}_3\!\times\!\mathbf{v}_1\right)^z.
    \label{eq:chirality}
\end{equation}
This takes the maximal value of $\kappa = \pm 4$ if the vectors are aligned in a $120^\circ$ order in the $x$-$y$ plane (for vectors of length $|\mathbf{v}_i|= 2$) and $\kappa = 0$ for collinear or randomly oriented vectors. 
The overall sign will be fixed by the sign of $J^\mathrm{DM}_{ij}$, which lifts the degeneracy between states of positive and negative chirality.
Fig.~\ref{fig:classical_observables}~(b) shows the ground-state chiralities
\begin{equation}
\begin{aligned}
    \kappa^\tau &\equiv \kappa(\langle\btau\rangle) \\
    \kappa^{\sigma\tau} &\equiv \kappa(\langle\sigma^x\btau\rangle) + \kappa(\langle\sigma^y\btau\rangle) + \kappa(\langle\sigma^z\btau\rangle)
\end{aligned}
\label{eq:specific-chiralities}
\end{equation}
as a function of $\Delta_V$. For $\Delta_V < -91 \mev$ the ground state clearly shows IVC $120^\circ$ order in $\bsigma\btaup$ with $\kappa^{\sigma\tau} = +4$. Explicitly, such a state can be obtained by starting, e.g., from $|\sigma^x\tau^x\rangle$ and then applying rotations in the valley $x$-$y$ plane as
\begin{equation}
    \label{eq:ivc-120}
    |\Psi_a\rangle^\mathrm{IVC}_{120^\circ} = e^{-i \tau^z \theta^a} |\sigma^x\tau^x\rangle \,,
\end{equation} 
with the angles $\theta^a = (0, 2\pi/3, 4\pi/3)$ for the three sublattices of the $120^\circ$ order. As before, eigenstates of $\btaup$ also lie in the ground-state manifold, but thermal fluctuations only select eigenstates of $\bsigma\btaup$.

Close to the Mott transition at $\Delta_V = -20 \mev$  we also observe finite ground-state chiralities, this time in both $\btaup$ and $\bsigma\btaup$ and with a negative sign. They do not, however, take fixed values but decrease with increasing $\Delta_V$, indicating that some form of incommensurate (ICS) order is likely realized. ICS order is, by its nature, difficult to quantify and visualize in real space. 
To remedy this, we calculate the spin-valley spin-valley correlation functions $\langle \bsigma_i \bsigma_j\rangle$, $\langle \boldsymbol{\tau}^\perp_i \, \boldsymbol{\tau}^\perp_j\rangle$ and  $\langle \bsigma_i\boldsymbol{\tau}^\perp_i \, \bsigma_j\boldsymbol{\tau}^\perp_j\rangle$ and obtain the corresponding structure factors via a straight-forward Fourier transform%
\footnote{Since we consider finite lattice sizes one has to be careful to only consider momenta allowed by the periodic boundary conditions to avoid unphysical artifacts in the structure factors.}%
, as depicted in Fig.~\ref{fig:classical_sfs_field}. The defining feature of the ICS states it then the position $\mathbf{k}_\mathrm{max}$ where the structure factors have their maxima, whose absolute values are shown in  Fig.~\ref{fig:classical_observables}~(c). 

In the ferromagnetic region peaks in the structure factor appear at the $\Gamma$ point and for $120^\circ$ order at the $K$ and $K^\prime$ points, as expected. In the ICS phase the ordering vectors lie at incommensurate momenta, which move between the $\Gamma$ and $K$ points for $\btaup$ and $\bsigma\btaup$ and between the $\Gamma$ and $M$ point for $\bsigma$. Only in a small region the spin structure factor has a peak exactly at the $M$ point, which indicates stripe order.

The energy of the FM and $120^\circ$ ordered states, as defined above, can be  calculated exactly and compared to the ground-state energy from our numerical minimization, showing very good agreement. At the phase boundary between the IVC $120^\circ$ and IVC FM states their energies cross, creating a sharp kink in the ground-state energy, which suggests a first-order transition. All other $T = 0$ transitions appear continuous (see appendix~\ref{app:mc} for supplemental data).

\subsection{Thermodynamics of the spin-valley model}

\begin{figure*}
    \centering
    \includegraphics{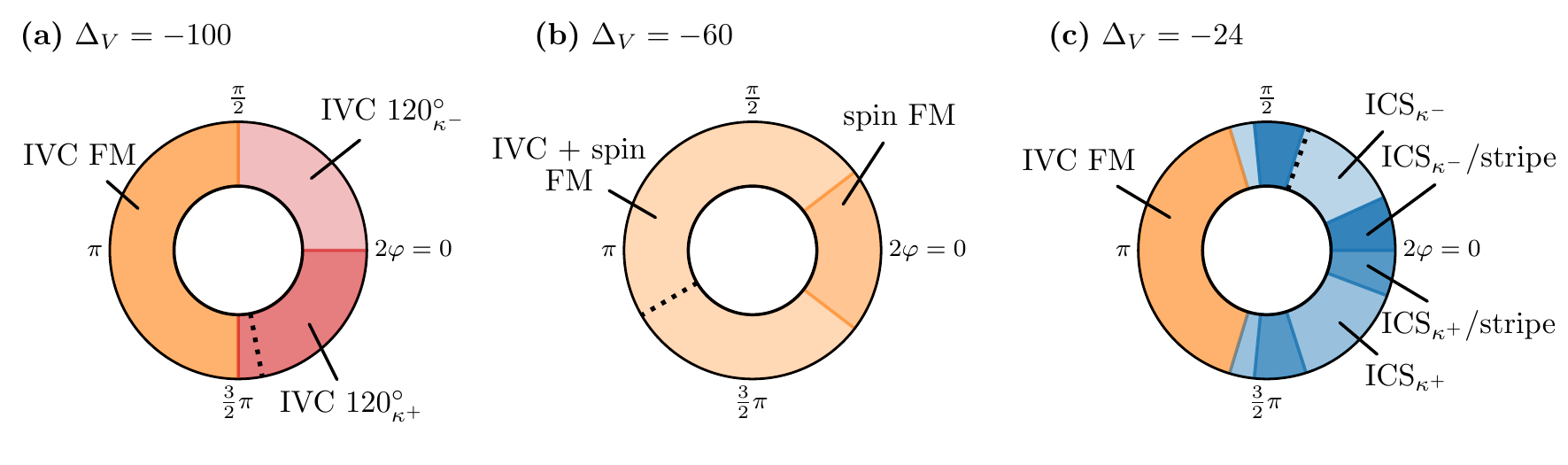}
    \includegraphics{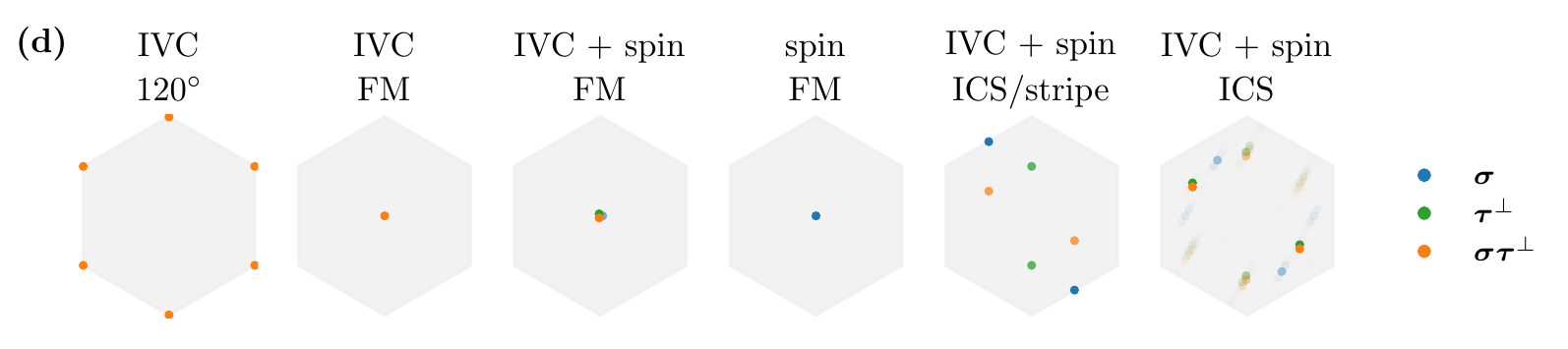}
    \caption{{\bf Semi-classical phase diagram as a function of the phase}. \textbf{(a)-(c)} Phases and phase boundaries for $J_1$ and $J_2$ fixed to their TG/h-BN estimates at  $\Delta_V=-100, -60, -24$, but with varying phase $\varphi$. Varying the phase $\varphi$ effectively tunes the in-plane nearest neighbor valley couplings $J_{ij}^\mathrm{XXZ}$ and $J^\mathrm{DM}_{ij}$ which both break the SU(4) symmetry. The dotted black lines show the initial estimate for $\varphi$ at these $\Delta_V$. $\kappa^{\pm}$ denotes the sign of the chirality (when non-zero). All ICS phases are of IVC + spin type. \textbf{(d)} Ground-state (static) structure factors. The color and opacity depict the sector and the relative magnitude, respectively. Points that lie on top of each other are slightly shifted for visibility.}
    \label{fig:classical_phase_diagram_flux}
\end{figure*}

Let us close this discussion of the various ground-state orders with a brief overview of their finite-temperature stability. 
Summarized in Fig.~\ref{fig:thermodynamics-field} we show specific heat traces along with their energy as a function of temperature
for the three principal phases of our model, the $120^\circ$ ordered state for $\Delta_V=-100$~meV, 
the spin-valley ferromagnet for $\Delta_V=-60$~meV and the incommensurate spiral phase for  $\Delta_V=-24$~meV.
All three phases show a sharp peak-feature in the specific heat at their respective thermal phase transitions that scales/diverges with increasing system size. 
The nature of these transitions appears to be continuous for the $120^\circ$ and ferromagnetically ordered states, while the double-peak structure in the energy histogram and the associated latent heat jump in the energy at the respective transition temperature indicates a first-order transition for the incommensurate phase.
All three ordered states exhibit a low-temperature specific heat saturation of $c_V(T\rightarrow0) = 5$, indicating the presence of {\it ten} harmonic zero modes~\cite{Chalker-1992} in their ordered states. This extraordinarily large number of zero modes (compared to, say, an $O(3)$ ferromagnet with two harmonic zero modes) can be rationalized as follows: 
The local Hilbert space of an $\mathfrak{su}(4)$ spin-valley model at a filling of two partons per side is six-dimensional. In the semi-classical picture, the state on each site is, therefore, parametrized by a six-dimensional complex valued vector. Counting the real and imaginary part of each component, but excluding one parameter for the normalization and an arbitrary phase, this leaves exactly ten parameters per site, each of which contributes a harmonic zero mode in the ordered phases of our model.

\subsection{Role of the SU(4) symmetry breaking couplings}\label{sec:su4break}

%
In discussing the ground-state orders of the spin-valley model we have seen that in particular the SU(4) breaking couplings $J_{ij}^\mathrm{DM}$ and $J_{ij}^\mathrm{XXZ}$ as well as their relative magntitudes are essential in determining the type of order that we  observe.  We now want to explore the role of these SU(4) breaking couplings in a more general phase diagram of the Hamiltonian~\eqref{eq:hamiltonian} where we go beyond the somewhat fine-balanced parametrization following 
Zhang and Senthil's model building in Ref.~\cite{zhang2019}.
To this end, we fix the value of the potential difference $\Delta_V$ (and the corresponding values of $J_1$ and $J_2$) to one of the three principal ordered phases ($120^\circ$ order, FM and ICS phase), lock the Hund's coupling to $J_H = 0.136\mev$, and then scan the
the SU(4) breaking couplings over a broad range by calculating phase diagrams as a function of the phase~$\varphi \equiv |\varphi_{ij}|$ in $0 \leq 2\varphi \leq 2\pi$, which determines the values of $J_{ij}^\mathrm{DM}$ and $J_{ij}^\mathrm{XXZ}$  via Eq.~\eqref{eq:su4-breaking-couplings}. In the region around $2\varphi\approx \pi$ the coupling $J_{ij}^\mathrm{XXZ}$ is dominant and ferromagnetic and as $J_{ij}^\mathrm{DM}$ is small we expect FM ground states with vanishing chirality. Moving away from $2\varphi\approx \pi$ the absolute value of $|J_{ij}^\mathrm{XXZ}|$ decreases and becomes smaller than $|J_{ij}^\mathrm{DM}|$ for $2\varphi \lesssim \pi/2$ and $2\varphi \gtrsim 3\pi/2$, which may induce $120^\circ$ or ICS order with finite chirality in this region.

The resulting phase diagrams as a function of $\varphi$ are shown in Fig.~\ref{fig:classical_phase_diagram_flux}~(a)-(c), 
for the $120^\circ$ order ($\Delta_V = -100$~meV), FM  ($\Delta_V = -60$~meV)  and ICS phase  ($\Delta_V = -24$~meV). In each plot the value of $\varphi$ corresponding to the estimate for TG-hBN in Fig.~\ref{fig:tghbn-phase-diagram}~(a) is indicated by a dotted black line. 
The structure factors corresponding to the different phases are depicted in Fig.~\ref{fig:classical_phase_diagram_flux}~(d).

For $\Delta_V = -100 \mev$ with $J_1>0$ and $J_2\approx 0$ an IVC $120^\circ$ order with positive chirality was found for the original $\varphi$. 
This order is preserved for $3\pi/2 < 2\varphi < 0$. Changing the sign of the phase (i.e. going to $0 < 2|\varphi{ij}| < \pi/2$) simply changes the sign of the chirality and not the type of order. 
In the region $\pi/2 < 2\varphi< 3\pi/2$, however, instead of $120^\circ$ order we get an IVC FM, stemming from the large negative value of the coupling~$J_{ij}^\mathrm{XXZ}$.

For $\Delta_V = -60 \mev$ the dominant coupling is $J_1 < 0$ and $|J_2|$ is still very small, resulting in IVC + spin FM order which remains stable for a large region of $\varphi$. 
Only for a phase close to $\varphi = 0$ a FM appears that has a finite magnetization only in the spin with $|\langle \bsigma\rangle|=2$. Such a truly long-ranged state should, in principle, not be allowed at finite temperatures due to the Mermin-Wagner theorem \cite{Mermin-1966}. 
As such we deem the associated finite-temperature feature to be a finite-size induced thermal crossover.

In the IVC + spin ICS phase at $\Delta_V = -24 \mev$,  the original order is again preserved for a large region of $\varphi$, with a change in chirality when changing the sign of the phase. 
Around $2\varphi = 0, \pi/2, 3/2\pi$ the spin-structure factor shows ordering vectors at the $M$~point again implying spin stripe order which also appeared in~Fig.~\ref{fig:tghbn-phase-diagram}. 
For a flux around the value $2\varphi=\pi$, the system again transitions into the IVC FM order.

In summary, there is a strong tendency towards FM states for $2\varphi \in [\pi/2, 3\pi/2]$, where the $J^\mathrm{XXZ}_{ij}$ dominates and a tendency towards finite-chirality states for  $2\varphi \in [3\pi/2, \pi/2]$, where $J^\mathrm{DM}_{ij}$ dominates. 
If $J_1$ is negative and sufficiently large, FM states remain for all $\varphi$. 
If $J_1$ is positive, the type of chiral state is determined by the magnitude of $J_2$, where for small values IVC $120^\circ$ order is favored, and for larger $J_2\approx J_1>0$ IVC + spin ICS or stripe order emerges. 
We never observe valley polarization, but only IVC, IVC + spin, or spin order.


\section{Quantum model}
\label{sec:frg}

In the previous section we have shown that the Hamiltonian~\eqref{eq:hamiltonian} realizes a plethora of ordered states that can be well captured in a semi-classical  analysis. Now, we turn to the question of how quantum fluctuations alter the picture, with particular interest in finding regions where spin-valley order is destabilized in favor of a quantum disordered ground state.
An established method for finding these regimes is the pseudo-fermion functional renormalization group (pf-FRG), which has, by now, been applied to a variety of spin models to resolve the competition between ordered and disordered ground states in the presence of quantum fluctuations~\cite{PhysRevB.83.024402,PhysRevB.83.064416,PhysRevB.84.014417,PhysRevB.84.100406,PhysRevB.86.155127,PhysRevB.90.100405,PhysRevB.92.220404,PhysRevB.94.224403,PhysRevB.94.235138,PhysRevLett.120.187202,PhysRevB.104.L220408,PhysRevLett.120.057201,PhysRevX.9.011005}.
In a technical development, the pf-FRG approach  was recently extended (by some of us) ~\cite{kiese2020, gresista2022} from conventional $\mathfrak{su}(2)$ spin models \cite{woelfle2010} to the $\mathfrak{su}(4)$ spin-valley models of interest here. 
After a short introduction of the basic concepts of the pf-FRG method, we employ the approach to investigate the role of quantum fluctuations on the phase diagrams previously explored in the semi-classical analysis.

\subsection{Pseudo-fermion functional renormalization group}

\begin{figure}[t!]
    \centering
    \includegraphics{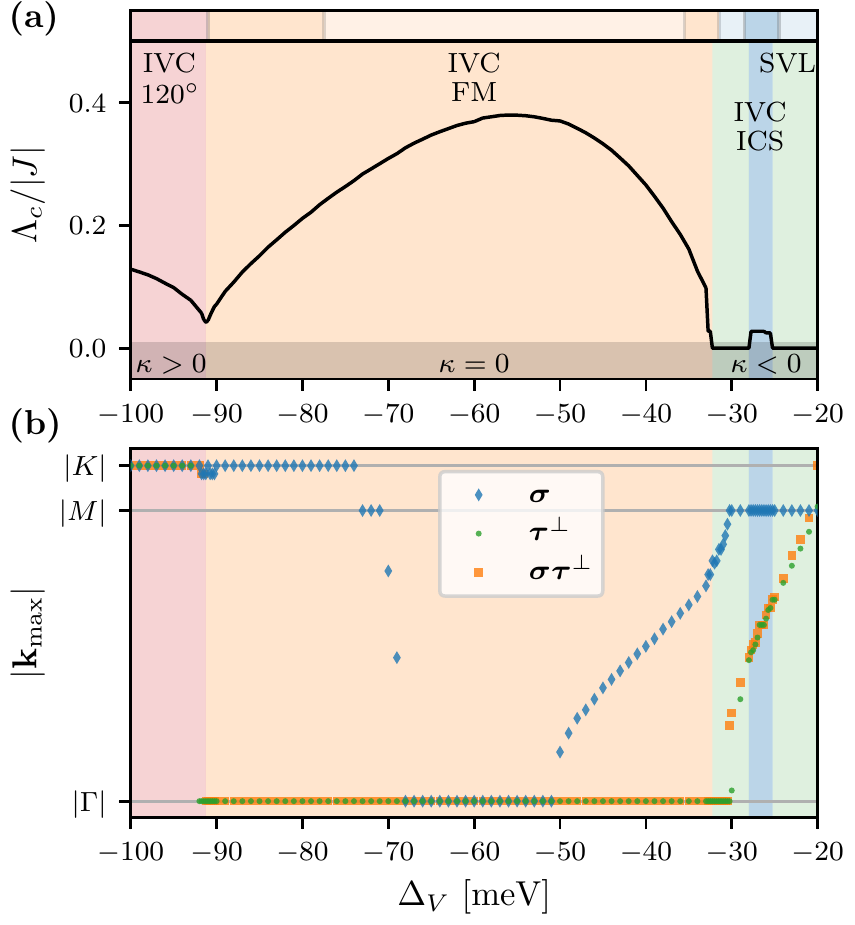}
    \includegraphics[width = \columnwidth]{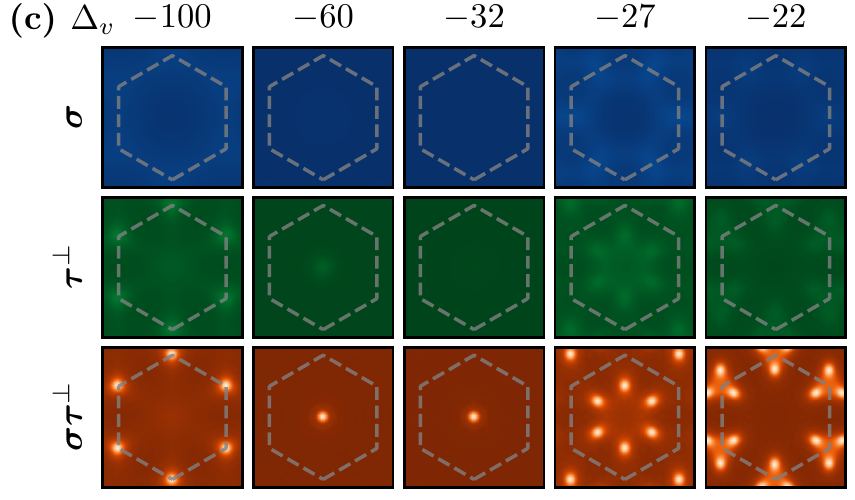}
    \caption{{\bf Quantum phase diagram of the TG/h-BN Hamiltonian} as a function of $\Delta_V$ from pf-FRG. \textbf{(a)}~Critical scale $\Lambda_c$ where the pf-FRG flow shows instabilities. The background color shows the phase boundaries and the small bar on top shows the semi-classical results for comparison. \textbf{(b)}~Magnitude of the ordering vector given by the momentum $\mathbf{k}_\mathrm{max}$ at which the structure factor is maximal. Note that we show all sectors, even though $\bsigma$ and $\btaup$ are very small compared to $\bsigma\btaup$. \textbf{(c)} Static structure factor at $\Lambda$ slightly above $\Lambda_c$. All sectors are plotted at the same color scale (for fixed $\Delta_V$). Compared to the semi-classical calculation (see Fig.~\ref{fig:tghbn-phase-diagram}), the spin-valley sector is always clearly dominant and no IVC + spin phases with order in multiple sectors appear, which stems from the fact the pf-FRG has difficulties to resolve such states (as explained in the main text).
    In the vicinity of the semi-classical ICS phase we observe regions with no flow breakdown, indicating a putative spin-valley liquid (SVL) ground state. The staggered vector chirality $\kappa$ changes sign consistent with the semi-classical result.}
    \label{fig:quantum_phase_diagram_field}
\end{figure}
\begin{figure*}
    \centering
    \includegraphics{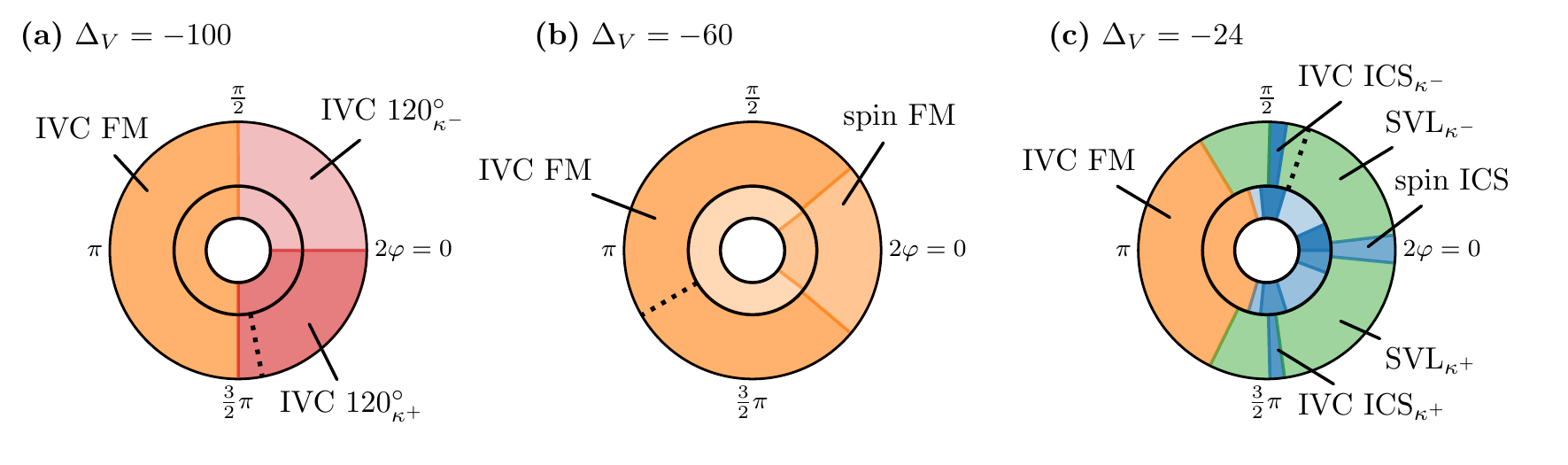}
    \caption{{\bf Quantum phase diagram as a function of the phase} from pf-FRG. The coupling $J_1$ and $J_2$ are fixed to their TG/h-BN estimates for different $\Delta_V$. The dotted black lines show the initial estimate for $\varphi$ at the corresponding $\Delta_V$. The inner circles show the semi-classical results (see Fig.~\ref{fig:classical_phase_diagram_flux} for comparison). The main discrepancy occurs in \textbf{(c)}. Here, the phase boundaries are slightly shifted and we observe disordered,  putative SVL liquid phases in the regime of the semi-classical ICS order. Around $\varphi = 0$ semi-classical ICS/stripe order turns into dominant ICS order in the spin. $\kappa^{\pm}$ denotes the sign of the chirality (when non-zero).}
    \label{fig:quantum_phase_diagram_flux}
\end{figure*}
The central idea of the FRG approach~\cite{WETTERICH199390} is to treat the system by not considering all energy scales at once, but starting from a known high-energy limit and iteratively including lower energy scales until the full theory is recovered~\cite{RevModPhys.84.299,Dupuis:2020fhh}. 
This is achieved by introducing an infrared cutoff $\Lambda$ into the theory so that at $\Lambda \to \infty$ all correlation functions are determined simply by the bare couplings in the Hamiltonian, and at $\Lambda = 0$ the full, physical correlation function reemerge. A derivative with respect to the cutoff $\Lambda$ generates an hierarchy of differential equations, the \emph{flow equations}, which govern the evolution of the correlation functions between these two limits.

The integration of the flow equations down to $\Lambda = 0$, however, is only possible if no indications of spontaneous symmetry breaking occur. The onset of such symmetry breaking is signaled by the divergence of a susceptibility, associated with an order parameter, when approaching a characteristic scale $\Lambda_c$, which we dub the \emph{critical scale}. 
To probe spin-valley orders with bilinear order parameters, we consider the Fourier transformed 
Green's function
\begin{equation}
    \label{eq:magnetic-susceptibility}
    \chi_{ij}^{\mu\nu\kappa\lambda\Lambda}(\omega) = \int_0^\infty\!\!\! d\tau e^{i\omega \tau}\! \left\langle \hat{T}_\tau (\sigma^\mu_i\tau_i^\kappa)(\tau)(\sigma^\nu_j\tau_j^\lambda)(0)\right\rangle^\Lambda\!\!,
\end{equation}
where $\hat{T}_{\tau}$ is the time-ordering operator in imaginary time $\tau$. 
Due to finite numerical accuracy the divergence is usually softened to a cusp or peak. 
Below the critical scale the flow is no longer physical, and the numerical integration of the flow equations has to be stopped.
If such a \emph{flow breakdown} occurs, the dominant component of $\chi_{ij}^{\mu \nu \kappa \lambda}$ tells us in which sector the system wants to order, whereas its Fourier transform (the momentum dependent \textit{structure factor}) allows us to determine the type of order. 
The absence of a flow breakdown, on the other hand, leaves open the possibility of a spin-valley liquid or a different quantum disordered ground state.

The symmetries of the Hamiltonian are preserved in the flow equations and carry over to the spin-valley spin-valley correlations. 
To distinguish the different types of order in the sectors defined in Eq.~\eqref{eq:sectors}, consider the diagonal components
\begin{equation}
\begin{aligned}
    &\chi_{ij}^\bsigma = \chi_{ij}^{\mu\mu dd} \sim \langle \bsigma_i \bsigma_j \rangle 
    \,,\\
    &\chi_{ij}^{\btaup} = \chi_{ij}^{ddxx} = \chi_{ij}^{ddyy} \sim \langle \boldsymbol{\tau}_i^\perp \boldsymbol{\tau}_j^\perp \rangle\,, \\
    &\chi_{ij}^{\bsigma\btaup}= \chi_{ij}^{\mu\mu xx} = \chi_{ij}^{\mu\mu yy} \sim \langle \bsigma_i^{}\boldsymbol{\tau}_i^\perp \bsigma^{}_j\boldsymbol{\tau}_j^\perp \rangle\,, \\
    &\chi_{ij}^{\tau^z} = \chi_{ij}^{ddzz} \sim \langle \tau^z_i \tau^z_j \rangle\,, \\
    &\chi_{ij}^{\bsigma\tau^z} = \chi_{ij}^{\mu\mu zz} \sim \langle \bsigma^{}_i\tau^z_i \bsigma^{}_j\tau^z_j \rangle\,,
\end{aligned}
\end{equation}
with $\mu = x, y, z$ (no summation). These are the same sectors as defined in the previous section for the semi-classical approach and, in principle, we can detect the same types of semi-classical order. At a flow breakdown, however, the subdominant components of the correlations are usually suppressed and only the dominant component diverges. This makes it difficult to resolve IVC + spin states with simultaneous order in multiple sectors. The valley out-of-plane sectors $\chi_{ij}^{\tau^z}$ and $\chi_{ij}^{\bsigma\tau^z}$ are always negligibly small and are not considered in the following. The off-diagonal components $\chi^{ddxy} = -\chi^{ddyx} \sim \langle\tau^x \tau^y\rangle$ and $\chi^{\mu\mu xy} = -\chi^{\mu\mu yx} \sim \langle\bsigma\tau^x \bsigma\tau^y\rangle$ are finite but also small. They can, however, still be used to determine a sign change in the staggered vector chirality defined in Eqs.~\eqref{eq:chirality} and~\eqref{eq:specific-chiralities}, see appendix~\ref{app:frg-data} for details.

\begin{figure}
    \centering
    \includegraphics{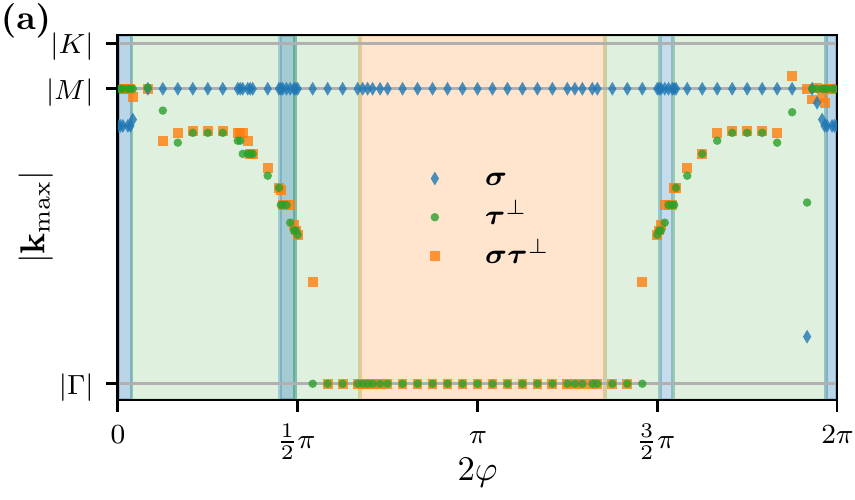}
    \includegraphics{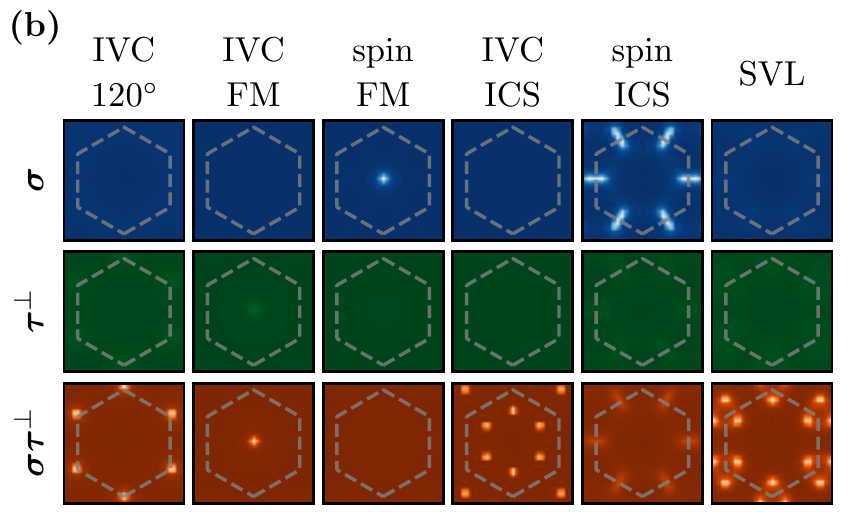}
    \caption{{\bf Quantum structure factors} from pf-FRG \textbf{(a)} Magnitude of the ordering vector given by the momentum $\mathbf{k}_\mathrm{max}$ where the structure factor is maximal for the phase diagram corresponding to Fig~\ref{fig:quantum_phase_diagram_flux}~(c) ($\Delta_V = -24$). 
    (b) Full static structure factors at $\Lambda$ slightly above $\Lambda_c$ for the phases found in Fig.~\ref{fig:quantum_phase_diagram_flux}. All sectors a plotted at the same color scale (in one phase).}
    \label{fig:quantum_sfs_flux}
\end{figure}
\begin{figure}
    \centering
    \includegraphics{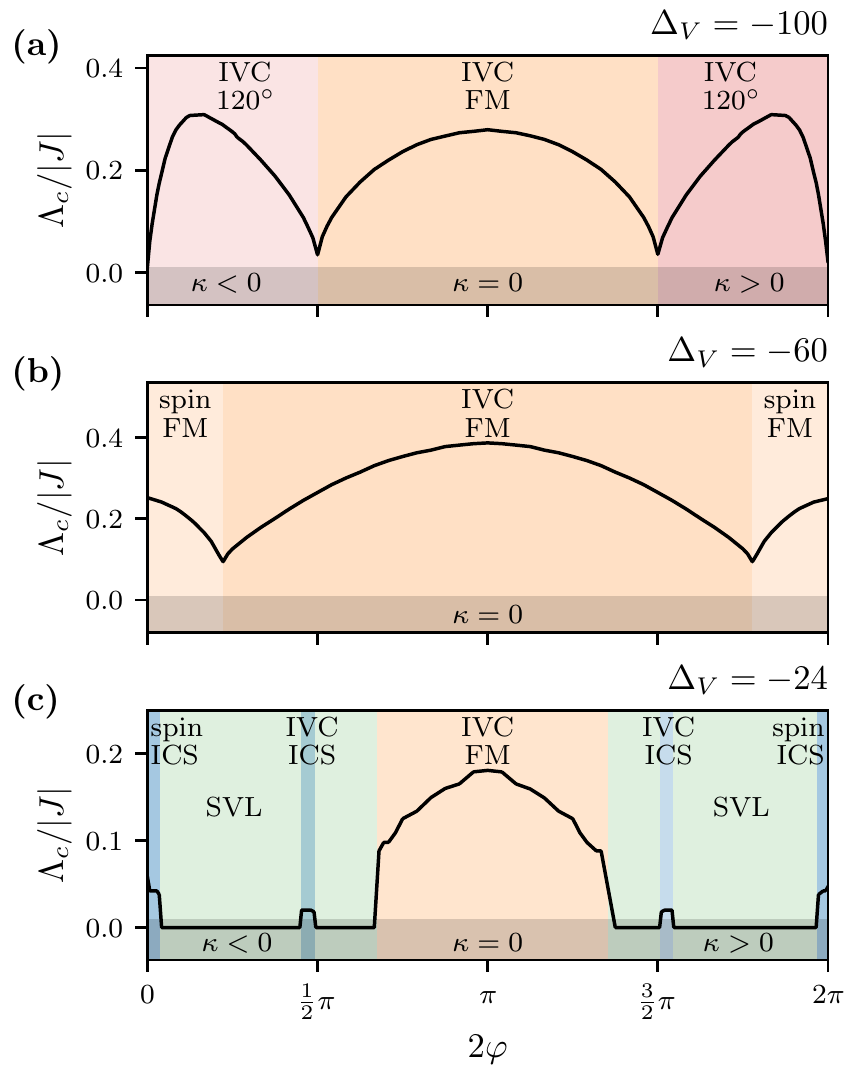}
    \caption{{\bf Critical scale as a function of the phase} for $J_1$ and $J_2$ fixed to the same parameters as in Fig.~\ref{fig:quantum_phase_diagram_flux}.  The black area highlights the value of $\Lambda/|J| =0.01$ below which we stop the numerical integration. The IVC ICS phases in \textbf{(c)} show flow break downs very close to this value, at which the numerics become increasingly unreliable.}
    \label{fig:quantum_thermodynamics_flux}
\end{figure}
%

\subsection{Quantum phase diagram for TG/h-BN spin-valley model}

We employ the pf-FRG to calculate the quantum counter parts to the semi-classical phase diagrams obtained in the previous section. Starting with the TG/h-BN inspired parameters, Fig.~\ref{fig:quantum_phase_diagram_field}~(a) shows the critical scale as a function of the potential difference $\Delta_V$ with labels denoting the types of ground state. 
The horizontal bar on top of the figure shows the semi-classical phase boundaries obtained in the previous section for comparison.  Fig.~\ref{fig:quantum_phase_diagram_field}~(b) shows the momentum $\mathbf{k}_\mathrm{max}$ for which the structure factor is maximal and (c) shows the full structure factor for several $\Delta_V$. 

Below $\Delta_V \lesssim -31 \mev$ the critical scale $\Lambda_c$ shows qualitatively similar behavior to the semi-classical transition temperature.  The structure factors imply IVC $120^\circ$ order with positive chirality for $\Delta_V \lesssim -91 \mev$ and a large FM phase centered around $\Delta_V \approx 60 \mev$. The position of the phase boundary between these phases matches the semi-classical result almost perfectly. In the FM phase, however, the pf-FRG does not clearly distinguish between the IVC and IVC + spin order, as already explained above. Instead, it always shows dominant $\chi^{\bsigma\btaup}$ implying an IVC FM ground state. The position of the maximum $\mathbf{k}_\mathrm{max}$ in the spin structure factor $\chi^\bsigma$ may indicate a crossover between different types of order, see Fig.~\ref{fig:quantum_phase_diagram_field}~(b), but the magnitude of the peaks relative to $\chi^{\bsigma\btaup}$ does not change significantly, see Fig.~\ref{fig:quantum_phase_diagram_field}~(c).

 For $\Delta_V \gtrsim -31 \mev$ the pf-FRG structure factors show peaks at incommensurate momenta, with dominant $\chi^{\bsigma\btaup}$ and negative chirality. 
 This is consistent with the semi-classically observed ICS order.  
 The flow of the structure factor, however, shows no flow breakdown down to the lowest considered scale indicated by the gray area in the figure. 
 This implies that no spin-valley order is present even for very low energy scales and suggests a putative spin-valley liquid ground state. 
 It is also possible that the system orders at scales too low for our numerical resolution. 
 This may explain the flow breakdowns in the small region close to the semi-classical ICS/stripe phase with a small critical scale of $\Lambda_c/|J| \approx 0.025$.
 %

\subsection{Role of the SU(4) symmetry breaking couplings}

We now turn to the phase diagrams with fixed $J_1$ and $J_2$ and varying phase $\varphi$.
Fig.~\ref{fig:quantum_phase_diagram_flux} shows the phases and their boundaries found by the pf-FRG in comparison to the semi-classical results. 
For $J_1$ and $J_2$ fixed to their values at $\Delta_V = -100 \mev$ and $\Delta_V = -60 \mev$ the phase diagrams agree remarkably well. 
The only significant difference is that semi-classical IVC + spin order again turns to only IVC order in the pf-FRG. 

For $\Delta_V = -24 \mev$, however, an extended putative SVL phase emerges where the semi-classical approach showed ICS order. 
The phase boundaries are also shifted notably. 
The SVL phase is interspersed by incommensurate order in the vicinity of the semi-classical  ICS/stripe phase. Close to $2\varphi = \pi/2, 3\pi/2$ we observe IVC ICS order and again a very small spin structure factor. 
Around $\varphi = 0$, on the other hand, the spin is dominant and shows ICS instead of stripe order. 
The corresponding structure factors and the position of their maxima are shown in Fig.~\ref{fig:quantum_sfs_flux}. The spin structure factor shows very faint stripe order everywhere expect close to $\varphi = 0 (2\pi)$. 
This may indicate that the corresponding stripe order observed in the semi-classical approach is an artifact of a finite lattice with periodic boundary conditions.

The phase boundaries are determined by the location of the minima in the critical scale, which is shown in Fig.~\ref{fig:quantum_thermodynamics_flux}. The critical scale in all ICS phases is again very close to the smallest scale we can reliably calculate and the results have to be interpreted with some caution due to increasing numerical uncertainties. It is both possible that at such low scales the flow breakdowns are numerical artifacts, or that the ICS phases are indeed larger, but order below $\Lambda_c/|J| = 0.01$. 


\section{Discussion}
\label{sec:discussion}

Motivated by the recent success in the synthesis and exploration of strongly-correlated moir\'e materials, we have  conducted a systematic study of triangular-lattice spin-valley models with \svsym~symmetry, which had been constructed as effective models for several moir\'e heterostructures.
While an approximate SU(4)~symmetry can be considered to be a good starting point for the study of their correlated phase diagram, SU(4)~breaking interactions such as XXZ anisotropies, Dzyaloshinskii-Moriya exchange, and on-site Hund's couplings, are expected to have a strong impact on magnetic and non-magnetic ground states.
Focussing on the case with a filling of two electrons (or holes) per moir\'e unit cell, we explored the \svsym~spin-valley model in a broad regime of coupling constants using semi-classical Monte Carlo simulations as well as complementary pseudo-fermion FRG calculations.
As the parameter space of the \svsym~model is large, we use material- and tuning-parameter specific predictions for TG/h-BN as a starting point and then generalize to a broader range by additional tuning of the XXZ anisotropies and Dzyaloshinskii-Moriya exchange couplings.

With both approaches, we consistently found a rich variety of ordered phases, including (anti-)ferromagnetic, incommensurate, and stripe order, which appear in the different regions of the model's parameter space. 
These states should therefore be considered as potential candidates for correlated insulator states appearing in moir\'e heterostructures described by the \svsym~spin-valley model.
We note that it is an ongoing discussion whether, e.g., the correlated insulating states in TG/h-BN are the result of a Stoner instability~\cite{Zhou_2021}, of the relevance of Mott correlations~\cite{calderon2022}, or of actual strong coupling~\cite{patri2022strong}.
Additional tuning or another materials composition may tip the scale into one or the other direction and our study can therefore shed some light at least on the strong-coupling side.

In certain parameter regimes it can be expected that quantum fluctuations  play an important role, which is not be covered by the semi-classical Monte-Carlo simluations.
In that case, the pseudo-fermion functional renormalization group is better suited to identify whether quantum disordered ground states are likely to occur.
Indeed, we found a broad parameter range where such spin-valley liquids may emerge, even in the presence of sizable SU(4)~breaking couplings.
We therefore conclude that strongly-correlated moir\'e heterostructures are a promising candidate materials to continue the search for and exploration of spin liquid physics.


\begin{acknowledgments}
We thank C. Hickey, J. Park, and A. Rosch for insightful discussions.
We acknowledge partial funding from the Deutsche Forschungsgemeinschaft (DFG, German Research Foundation) within Project-ID 277146847, SFB 1238 (project C02). 
The numerical simulations were performed on the Noctua 2 cluster at the Paderborn Center for Parallel Computing (PC2) and the CHEOPS cluster at RRZK Cologne.
MMS acknowledges support by the DFG through the DFG Heisenberg programme (Project-ID 452976698).
\end{acknowledgments}


\bibliography{main}

\clearpage 
\appendix

\section{Semi-classical Monte Carlo simulations}
\label{app:mc}

In this section we give additional details on the implementation of the semi-classical Monte Carlo calculations, discussing the definition of the semi-classical limit, the local Metropolis update used in our Monte Carlo algorithms, the finite-temperature calculations and the minimization procedure to obtain the ground state. 
We then present supplemental numerical data relevant for the results discussed in Sec.~\ref{sec:semiclass}, i.e. ground state energies obtained from the numerical minimization as a function of $\Delta_V$ and $\varphi$, the evolution of the latent heat of the first-order transition into ICS/stripe, the transition temperature as a function of $\varphi$, and specific heat saturation for $T \to 0$.

\subsection{Implementation}

\subsubsection*{Semi-classical limit}
As described in the main text, we define the semi-classical limit solely by the fact that there is no entanglement between two different lattice sites. This is enforced by considering only product states of the form 
\begin{equation}
\label{eq:product-state}
    |\psi\rangle = \otimes_{i} |\psi_i\rangle \,,
\end{equation}
where $|\psi_i\rangle$ is an arbitrary state in the local six-dimensional Hilbert space parametrized as
\begin{equation}
\label{eq:local-state}
    |\psi_i\rangle = \sum_{j=1}^6 b_i^j|\gamma^j\rangle\,,
\end{equation}
with normalized, six-dimensional complex vectors $|\mathbf{b_i}| = 1$ and a basis of the local Hilbert space $\{|\gamma^j\rangle\}$. 

Using the notation $|s_1 l_1, s_2 l_2\rangle$, where $s_{1/2}$ and $l_{1/2}$ are the spin and valley of the first and second electron, we choose the six basis states
\begin{equation}
\begin{aligned}
    |\gamma^j\rangle \in \big\{
    &\left|\uparrow+, \downarrow +\right\rangle,
     \left|\uparrow+, \downarrow -\right\rangle,
     \left|\downarrow+, \uparrow -\right\rangle,\\
    &\left|\uparrow+, \uparrow -\right\rangle,
     \left|\downarrow+, \downarrow -\right\rangle,
     \left|\uparrow-, \downarrow -\right\rangle 
    \big\}\,.
\end{aligned}
\end{equation}
For a lattice with $N$ sites and subtracting the normalization as well as an arbitrary local phase, the full state $|\psi \rangle$ is, therefore, parameterized by $N \cdot (12-2) = 10 N$ real numbers. 

The semi-classical energy, defined as
\begin{equation}
    H_\mathrm{sc}(\{\mathbf{b}_{i}\}) = \langle\psi| H |\psi\rangle\,,
\end{equation}
then is a function of these real numbers. To obtain finite temperature obseravbles, we follow the approach in Refs.~\cite{stoudenmire2009, hickey2014} and approximate the partition function as
\begin{equation}
Z = \int \prod_i \mathrm{d}\mathbf{b}_i \langle \Psi|e^{-\beta H}|\Psi\rangle \approx \int \prod_i \mathrm{d}\mathbf{b}_i\ e^{-\beta \langle \psi|H|\psi\rangle}.
\end{equation}
This is equivalent to a cumulant expansion to first order, which becomes exact in the limits of low $T \to 0$ and high temperature $T \to \infty$. Thermal expectation values can now be efficiently calculated using a Markov Chain Monte Carlo algorithm, as we describe in the following. 

\subsubsection*{Metropolis updates in spin-valley space}

To calculate finite-temperature observables in the semi-classical limit, we sample spin-valley configurations according to the Boltzmann distribution $\sim \exp(-\beta \langle\psi| H |\psi\rangle)$ using the Metropolis algorithm~\cite{LandauBinder} with local updates. 
Instead of a conventional spin configuration, however, we need to uniformly sample from the space of product states parameterized according to Eqs.~\eqref{eq:product-state},~\eqref{eq:local-state} by a 6-dimensional complex vector $\mathbf{b}_i$ on each site $i$. 
These vectors are normalized and can therefore be understood to live on a real 11-dimensional hypersphere parameterized by the real and imaginary part of each vector component. 
To uniformly sample on an $n$-dimensional hypersphere ($n$-sphere), we can use the method from Ref.~\cite{muller1959} and draw $n+1$ normally distributed random numbers and normalize the resulting vector afterwards. 
 A new state can, consequently, be generated by randomly selecting a site $i$ and then sampling a new local state
\begin{equation}
    \mathbf{b}_i^\prime = \frac{\mathbf{\Gamma}}{|\mathbf{\Gamma}|},
\end{equation}
where $\mathbf{\Gamma}$ is a six-dimensional complex vector, with the real and imaginary part of each component sampled from a normal distribution.
Such sampling on the full sphere, however, leads to very low acceptance rates  for low temperatures and in turn to very slow convergence of the results. To combat this, we generalize the update procedure proposed for classical $\mathfrak{su}(2)$ spins in Ref.~\cite{cardona2019} and utilize the \emph{Gaussian trial move}, which generates a new state in the `vicinity' of the original as
\begin{equation}
    \mathbf{b}_i^\prime = \frac{\mathbf{b}_i + \sigma_g \mathbf{\Gamma}}{|\mathbf{b}_i + \sigma_g \mathbf{\Gamma}|}.
\end{equation}
This is also an unbiased way of sampling the local Hilbert space, but with the benefit that the acceptance rate can be adjusted by
controlling the value of $\sigma_g$. Starting with a large $\sigma_g$ and then updating $\sigma_g$ after every tenth sweep according to 
\begin{equation}
\label{eq:update-cone-width}
    \sigma_g \to \frac{0.5}{1-R} \sigma_g,
\end{equation}
where $R$ is the acceptance rate during the last ten sweeps, this very quickly tunes the overall acceptance rate to approximately 50~\% and we observe a significant speedup in the convergence for low temperatures. 

\subsubsection*{Finite-temperature calculations}

A full finite-temperature Monte Carlo run for a temperature $T$ is divided in a thermalization phase for $N_t$ sweeps and measurement phase for $N_m = 10 N_t $ sweeps and proceeds as follows: 
For the thermalization phase, we start with a large temperature $T_i = 2|J|$ and $\sigma_g = 60$ and perform Monte Carlo sweeps using the update procedure explained above. 
For the first $\frac{3}{4} N_t$ sweeps the temperature is gradually lowered to the desired $T$ by multiplying it with the factor $ (T/T_i)^\frac{4}{3 N_t}$ after each sweep. 
For the remaining $\frac{1}{4} N_t$ sweeps the temperature is kept constant.  

During the measurement phase, temperature and $\sigma_g$ are kept constant for $N_m$ sweeps and observables are measured every tenth sweep. 
The statistical evaluation of measurements is done with the \texttt{BinningAnalysis} Julia Package~\cite{binninganalysis}.

Typical Monte Carlo runs in our simulations use a setup of up to $N=36^2$ lattice sites with periodic boundary conditions and $N_m = 4\cdot10^6$ sweeps per temperature, or up to $N_m = 20\cdot 10^6$ sweeps close to the transition temperature. 
\subsubsection*{Minimization to the ground state}

In order to obtain semi-classical ground-state observables, we first use simulated annealing to get close to the global energy minimum while avoiding possible local minima. We follow this by a stochastic gradient descent, which monotonically further lowers the energy towards the ground-state value.

Similar to the thermalization phase of a conventional Monte Carlo run, simulated annealing constitutes of performing Monte Carlo sweeps using the Metropolis updates described above while gradually lowering the temperature. We initialize the system at $T_i = 2|J|$ and $\sigma_g = 60$ and perform Metropolis updates for 4000 sweeps, or until $400N$ updates have been accepted (where $N$ is the number of sites), whichever comes first. Afterwards, we calculate the acceptance rate $R$ at the current temperature, adjust $\sigma_g$ according to Eq.~\eqref{eq:update-cone-width} and lower the temperature by 2~\%. When $\sigma_g$ has reached the minimal value of $\sigma_g = 0.05$ we keep it constant. We continue lowering the temperature until the acceptance rate is below $R_\mathrm{min}=0.001 \%$, after which we stop the calculation.

Starting from the so obtained state, we perform optimization sweeps using stochastic gradient descent. 
Here, the idea is to randomly pick a site $i$, obtain the energy $H_\mathrm{sc}^i$ as a function of only $\mathbf{b}_i$ (with $\mathbf{b}_{j\neq i}$ fixed), and then minimize this energy using gradient descent. 
To preserve the normalization $|\mathbf{b}_i| = 1$, however, $H_\mathrm{sc}^i$ needs to be minimized on the 11-sphere spanned by the real and imaginary part of $\mathbf{b}_i$. 
To this end, we calculate the gradient $\nabla H_\mathrm{sc}^i$ on the sphere using the finite differences backend of the \texttt{Manifolds} Julia package~\cite{manifolds}, and then perform gradient descent using the \texttt{Manopt} Julia package \cite{manopt}. 
The gradient descent on a single site is stopped when the norm of the gradient is below $0.001 |J|$. 
Performing this for $N$ randomly chosen sites constitutes one optimization sweep. 
We find that after $N_o = 60$ optimization sweeps the energy does not  significantly change anymore and we stop the minimization. 
From the resulting state we can then calculate ground-state observables.
To confirm that we are not stuck in a local energy minimum, we perform several of these minimizations and compare the resulting energies.

\begin{figure}
    \centering
\includegraphics{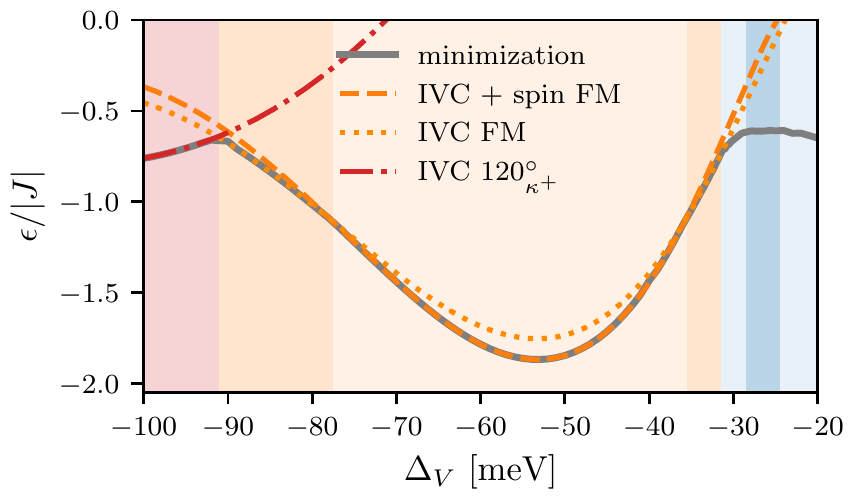}
    \caption{{\bf Ground-state energy per site for TG/h-BN inspired parameters.} The solid gray line shows the result of the numerical minimization. The dashed/dotted lines depict the exact energies of the FM and $120^\circ$ type states. The kink at the transition between FM and $120^\circ$ order suggests a first-order transition, while all other transitions appear to be continuous.}
    \label{fig:gs-energies-field}
\end{figure}

\subsection{Supplemental numerical data}
\label{app:mc-data}
\subsubsection*{Ground-state energies}
\begin{figure}
    \centering
\includegraphics{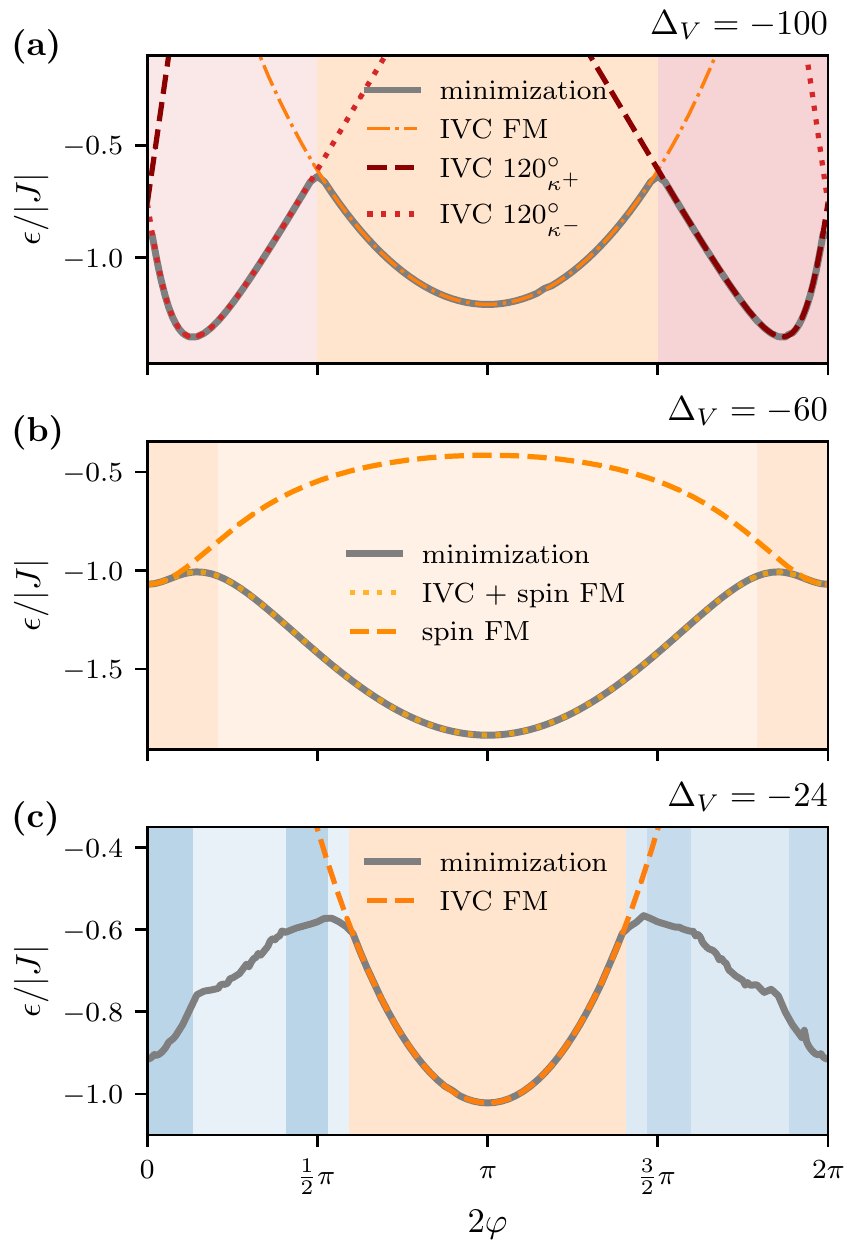}
    \caption{{\bf Ground-state energy per site as function of the phase} for $J_1$ and $J_2$ fixed to the same parameters as in Fig.~\ref{fig:classical_phase_diagram_flux}. The solid gray line shows the result of the numerical minimization. The dashed/dotted lines depict the exact energies of the FM and $120^\circ$ type states. The transition between the $120^\circ_{\kappa^+}$, $120^\circ_{\kappa^-}$ and FM order shows a clear kink in the energy suggesting a first-order transition, while transitions between the two FM phases appear continuous. The nature of the transition in the ICS phase in \textbf{(c)} (blue background) can not be conclusively determined from the numerical data.}
    \label{fig:gs-energies-flux}
\end{figure}
 In this section, we present the ground-state energies obtained from the numerical minimization to confirm the correct identification of the FM and $120^\circ$ states, and to elucidate the nature of the $T = 0$ transitions between the different ground-state orders.  Starting with the TG/h-BN inspired parameters, Fig.~\ref{fig:gs-energies-field} shows the energy from the minimization compared to the exact energies of the IVC FM (e.g. given by $|\sigma^x\tau^x\rangle$), the IVC + spin FM (defined in Eq.~\eqref{eq:mixed-state}), and the IVC $120^\circ$ state (defined in Eq.~\eqref{eq:ivc-120}). The numerical minimization almost exactly matches the energy of the lowest lying state, apart from the ICS phase where we did not obtain an analytic expression for the ground state. At $T = 0$ the energy is equivalent to the free energy, in which a kink (i.e. a discontinuity in the first derivative) implies a first-order transition. The transition between $120^\circ$ and FM order precisely shows this behavior, while all other transitions appear to be continuous. 
 
 Similar behavior is found when varying $\varphi$ for fixed $J_1$ and $J_2$ as shown in Fig.~\ref{fig:gs-energies-flux}, where the transition between $120^\circ$ and FM phases shows a clear kink. To determine the energy of the IVC + spin FM shown Fig.~\ref{fig:gs-energies-flux}~(b), we numerically obtain the optimal value for $\delta$, which is the coefficient of the pure spin eigenstate in the definition of the IVC + spin FM, so that the state has minimal energy. The value $\delta$ has its minimum of $\delta \approx 0.52$ at $2\varphi = \pi$ in the center of the IVC + spin FM phase and continuously grows when moving towards the spin FM phase. In the close vicinity of $\varphi = 0\ (2\pi)$, $\delta\to\infty$ is the optimal value, which is equivalent to the pure spin FM, suggesting a continuous transition. The phase boundary shown is in this case determined by the position of a dip in the transition temperature (see Fig.~\ref{fig:tc-flux}~(b) below).  The nature of the transition in the ICS phase in Fig.~\ref{fig:gs-energies-flux}~(c) (blue background) can not be conclusively determined from the numerical data.
 
\subsubsection*{Thermodynamics}

\begin{figure}
    \centering
    \includegraphics{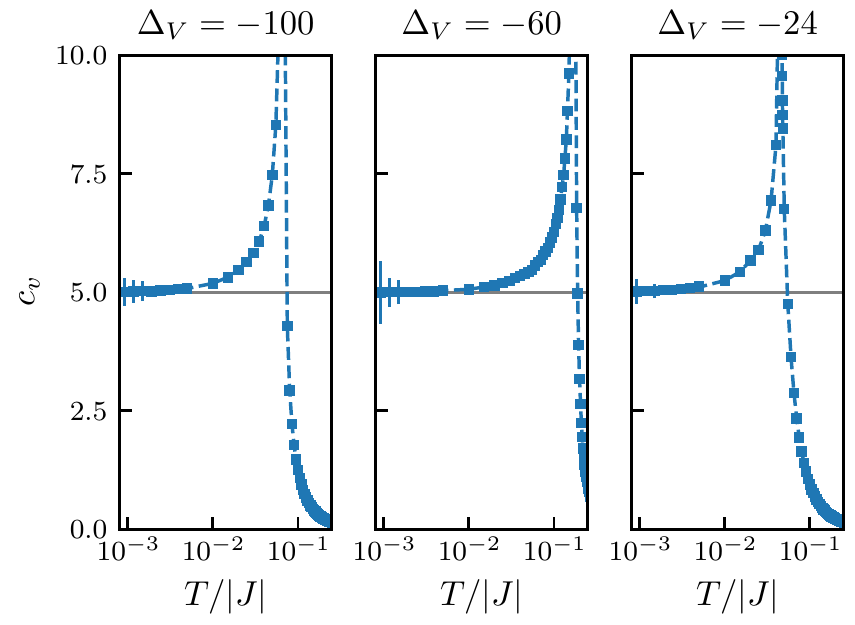}
    \caption{{\bf Specific heat saturation at low temperatures} for TG/h-BN inspired parameters deep in the $120^\circ$, FM and ICS phase and $L = 12$. We observe $c_v(T\to0) = 5$ for all types of order, also for those not explicitly shown here. This indicates the existence of ten harmonic zero modes.}
    \label{fig:heat_low_t}
\end{figure}

In Sec.~\ref{sec:semiclass} we have discussed the emergence of ten harmonic zero modes in all observed phases of our model, indicated by a specific heat saturating at $c_v(T \to 0) = 5$. Fig.~\ref{fig:heat_low_t} shows examples of the corresponding low temperature behavior of the specific heat in the three principal phases of our model. The other phases, not explicitly shown here, exhibit very similar low temperature behavior and we always find $c_v(T \to 0) = 5$.

\begin{figure}
    \centering
    \includegraphics{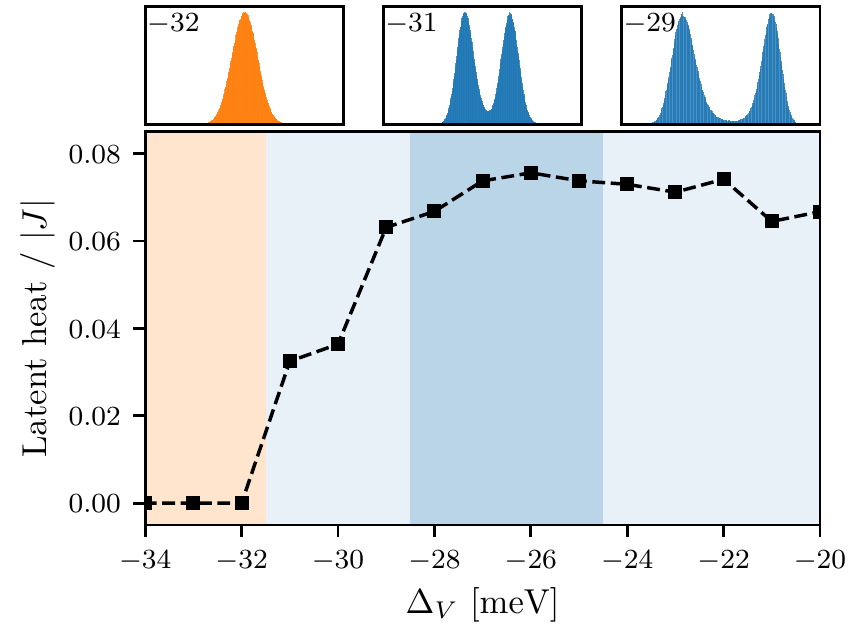}
    \caption{{\bf Latent heat} of the first-order transition into ICS/stripe order. Here the energy distribution at the transition temperature shows a double-peak structure, as shown in the top panels for $\Delta_V=-32, -31$ and $-29\mev$. The latent heat is given by the difference of the peak positions, which we determine by fitting double Gaussians to the energy distributions. This is done for a lattice size of $L=24$.}
    \label{fig:latent-heat-field}
\end{figure}

\begin{figure}
    \centering
    \includegraphics{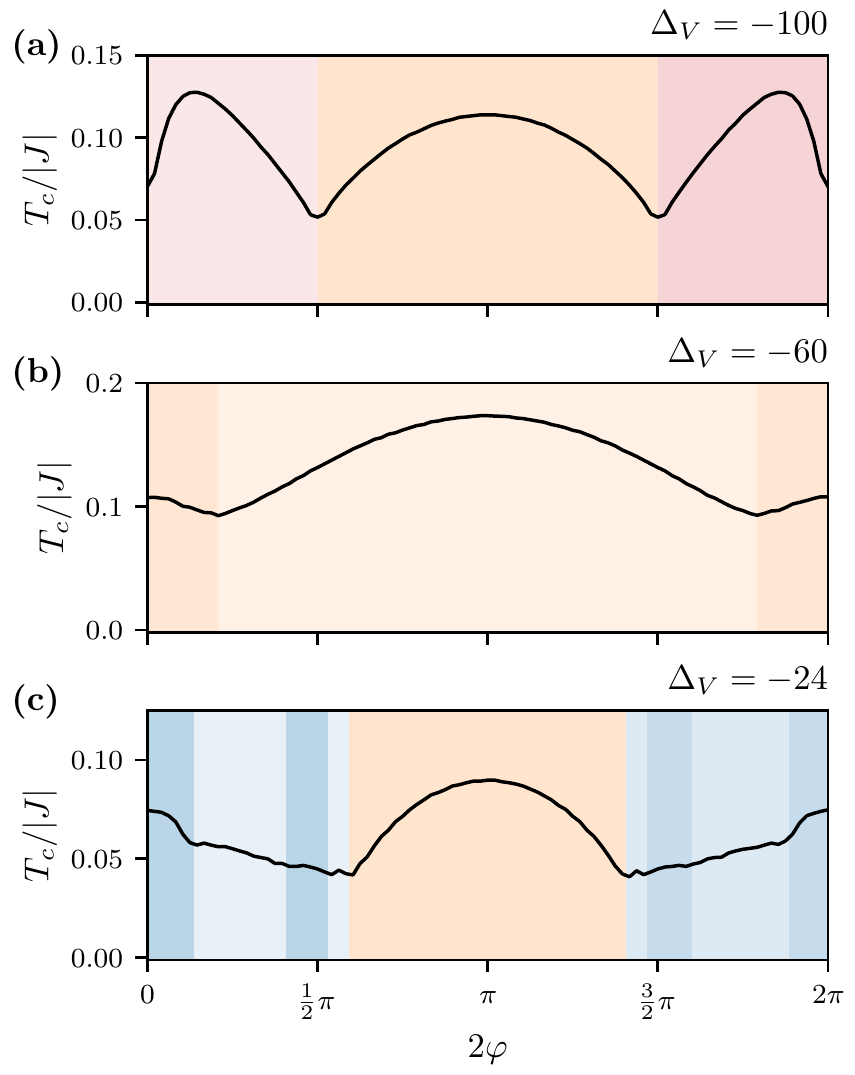}
    \caption{{\bf Transition temperature as a function of the phase} for $J_1$ and $J_2$ fixed to the same parameters as in Fig.~\ref{fig:classical_phase_diagram_flux}. $T_c$ is determined as the position of the maximum in the specific heat extrapolated to infinite lattice size.}
    \label{fig:tc-flux}
\end{figure}

\begin{figure*}
    \centering
    \includegraphics{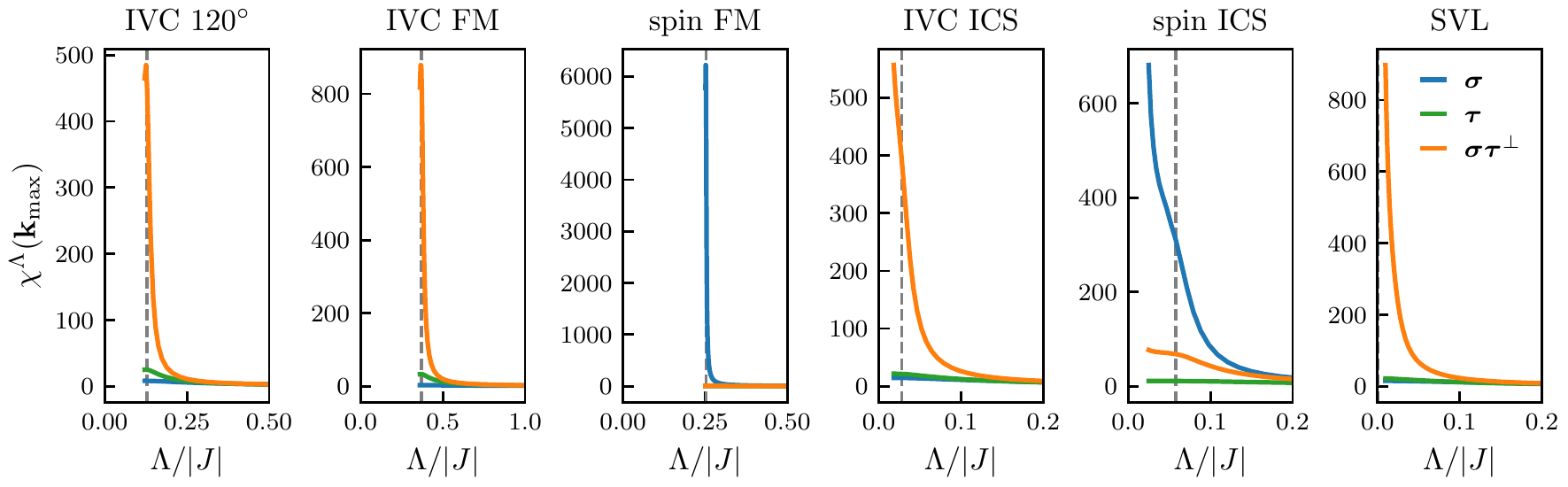}
    \caption{{\bf Renormalization group flow of the static structure factor} for the different phases of our model at the momentum of maximal intensity $\mathbf{k}_\mathrm{max}$. The dotted gray line shows the critical scale $\Lambda_c$ indicating the onset of magnetic order, which is determined by the position of maximal negative curvature. In the SVL phase the flow stays convex down to the lowest numerically computed cutoff $\Lambda/|J| = 0.01$, i.e. the curvature is always positive. This indicates the absence of magnetic order.}
    \label{fig:sf_flows}
\end{figure*}

At the transition into the ICS phase shown in the right column of Fig.~\ref{fig:thermodynamics-field} of the main text the energy as a function of $T$ shows a clearly visible discontinuity and the energy histogram a well-developed double-peak structure, both indicating a first order transition. Here we expand on the strength of this first-order transition by determining the latent heat released at this first-order transition. To this end, we fit double Gaussians to the energy histogram and calculate the difference of the two peak positions, which corresponds to the jump in the energy at the transition.  
The evolution of the so-determined latent heat is shown in Fig.~\ref{fig:latent-heat-field}. Deep inside the ICS/stripe ordered phase at $\Delta_V > -30$~ meV, the associated thermal phase transition exhibits a sizeable latent heat of about $0.07|J|$, indicating a strong first-order thermal transition. Upon approaching the transition to the ferromagnetically ordered phase, this latent heat quickly vanishes in a continuous way, indicating a softening of the transition into a continuous thermal transition as expected for a ferromagnet.

To conclude our discussion of the semi-classical thermodynamics, Fig.~\ref{fig:tc-flux} shows the transition temperature as a function of $\varphi$, using the same parameters as in Fig.~\ref{fig:classical_phase_diagram_flux}. For these plots $T_c$ is determined by the position of the maximum in the specific heat calculated for system sizes $L=12, 24, 36$ and then extrapolated to infinite lattice size using a linear fit. 

\section{pf-FRG simulations}
\label{app:frg-data}

We round off the manuscript by presenting additional details on the implementation of the pf-FRG and provide supplemental data for the corresponding calculations of the quantum model.

\subsection{Implementation}
The first step in the pf-FRG approach is to rewrite the spin-valley Hamiltonian by decomposing the spin-valley operators into complex pseudo-fermions as presented in Eq.~\eqref{eq:spin-valley-operators}.
It is important to note that the operators $f^{(\dagger)}_{isl}$ do {\it not} represent the original itinerant electrons of the underlying half-filled Hubbard model, but are in fact auxiliary degrees of freedom, which have to always fulfil the local number constraint of two pseudo-fermions per site, i.e. $n_i = 2$, cf. Eq.~\eqref{eq:half-filling}.
In the presence of particle-hole symmetry, this constraint is enforced on average, i.e. $\langle n_i \rangle = 2$, and our implementation maintains this symmetry\footnote{We note that exactly enforcing the constraint is notoriously difficult and has up to now only been achieved for $\mathfrak{su}(2)$ spin models using the \emph{Popov-Fedotov} trick~\cite{schneider2022} or a Majorona fermion representation~\cite{niggemann2021, niggemann2022}. 
Although quantitatively accurate results on small spin clusters could be obtained for temperatures $T / |J| \gtrsim 0.2$, their extrapolation to the zero temperature limit, considered here, has so far remained elusive within both approaches.}.
Particle-number fluctuations around the average are not expected to alter the qualitative behavior of physical observables obtained from the pf-FRG~\cite{kiese2020, kiese2022, thoenniss2020}.

The central idea of the FRG approach~\cite{WETTERICH199390} is to treat the resulting fermionic Hamiltonian by not considering all energy scales at once, but starting from a known high-energy limit and iteratively including lower energy scales until the full theory is recovered~\cite{RevModPhys.84.299,Dupuis:2020fhh}. 
This is achieved by introducing a regulator function $\Theta^\Lambda(\omega)$ in the bare propagator
\begin{equation}
    G_0(\omega) = {(i\omega)^{-1}} \to G_0^\Lambda(\omega) = \Theta^\Lambda(\omega) G_0(\omega),
\end{equation}
so that it satisfies the boundary conditions $G_0^{\Lambda\to\infty} = 0$ and $G_0^{\Lambda \to 0} = G_0$. 
A derivative with respect to the cutoff $\Lambda$ generates an infinite hierarchy of differential equations, the \emph{flow equations}, which govern the evolution of the one-particle irreducible \emph{vertex} functions. 
To become amenable to numerical computations the infinite hierarchy has to be truncated. 
Here, we choose the Katanin truncation~\cite{PhysRevB.70.115109}, which only considers the scale dependence of one- and two-particle correlation functions, {i.e.\
the frequency-dependent self-energy $\Sigma^\Lambda$ and the interaction vertex $\Gamma^\Lambda$.
This level of truncation} has been shown to successfully discriminate between ordered and disordered regimes~\cite{woelfle2010}.
The initial conditions of the vertex functions at $\Lambda \to \infty$ are given by the bare coupling constants in the Hamiltonian {and integrating out all fluctuations amounts to lowering the cutoff scale down to $\Lambda \to 0$.}

In our numerical implementation, which is based on the \texttt{ PFFRGSolver} Julia package \cite{kiese2020}, the continuous Matsubara frequencies are discretized in an adaptive grid with $N_\Sigma = 200$ frequencies for the one-particle vertex and $N_\Gamma = 40 \times 30 \times 60$ for the two-particle vertex,
where we checked for convergence with respect to increasing the number of frequencies. The infinite lattice is approximated by considering a real-space vertex truncation of $L=12$ lattice bonds, which effectively enforces a maximal correlation length. This is beneficial compared to periodic boundary conditions as incommensurate order is more easily resolved~\cite{PhysRevB.103.184407}. We integrate the flow equations,
 which for this setup amount to approximately $8 \cdot 10^7$ coupled differential equations, starting from a large cutoff $\Lambda/|J| = 20$ down to $\Lambda/|J| = 0.01$, below which the simulation starts to become unreliable. We label phases that show no flow breakdown above $\Lambda/|J| = 0.01$ as  putative spin-valley liquids (SVL) that exhibit a quantum disordered ground state. 

\subsection{Supplemental numerical data}

\subsubsection*{Renormalization group flow of the structure factor}

In Fig.~\ref{fig:sf_flows} we show the renormalization group flow of the static (equal-time) structure factor for the different phases of our model, calculated at the momentum $\mathbf{k}_\mathrm{max}$ with maximal intensity of the dominant sector. A divergence in the flow, which is usually softened to a kink or cusp by finite size effects, indicates the onset of magnetic order. We determine the associated critical scale $\Lambda_c$ as the $\Lambda$ where the structure factor has the maximal negative curvature (the second derivative). If the flow is convex down to $\Lambda/|J| = 0.01$, i.e. the curvature is always positive, we interpret the ground state to be a disordered, putative SVL state. In the case of a flow breakdown, we always observe either dominant $\chi^{\bsigma\btaup}$, indicating IVC order, or in $\chi^\bsigma$, indicating spin order. In contrast to the semi-classical model, the structure factor in the dominant sector is always close to a magnitude bigger than the others indicating that we never observe an IVC + spin state in the quantum model.
\begin{figure}[b]
    \centering
    \includegraphics{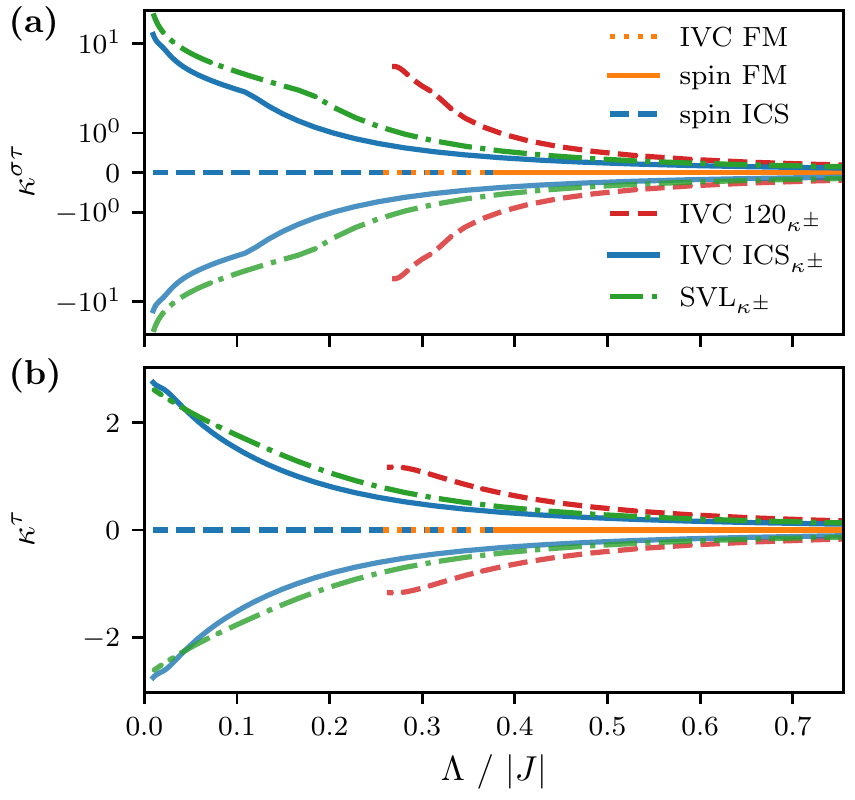}
    \caption{{\bf Renormalization group flow of the staggered chirality} deep inside the different phases of our model. FM states, or states that order only in the spin always show zero chirality. The chirality of all other states is consistent with the sign of the coupling $J_{ij}^\mathrm{DM}$ in Eq.~\eqref{eq:hamiltonian}.}
    \label{fig:quantum_chirality}
\end{figure}
%
\subsubsection*{Staggered vector chirality}
The sign of the staggered chirality, as defined in Eqs.~\eqref{eq:chirality}, \eqref{eq:specific-chiralities}, can also be calculated in the quantum model using the nearest neighbor correlations obtained from the pf-FRG flow. Since the flow equations preserve the symmetries of the Hamiltonian, which include a $C_3$ rotation and a sign change under inversion,  all nearest neighbor correlations along a triangle will be equal up to a a sign change between left and right pointing triangles. A sign change under inversion also implies $\chi^{\mu\mu xy}_{ij} = -\chi^{\mu\mu yx}_{ij}$. It follows that the chiralities are proportional to  $\kappa^{\tau} \sim \chi^{ddxy}_{\langle ij\rangle}$ and $\kappa^{\sigma\tau} \sim \chi^{\mu\mu xy}_{\langle ij \rangle}$, with $\langle ij\rangle$ being a nearest-neighbor bond in a right-pointing triangle. Fig.~\ref{fig:quantum_chirality} shows the resulting flow of the staggered chiralities in all phases of our model. The FM and pure spin orders show zero chirality. The chirality in all other phases exactly matches the sign changes of the $J^\mathrm{DM}_{ij}$ coupling of Eq.~\eqref{eq:hamiltonian}, as was the case for the semi-classical model. Even the SVL phase shows a finite chirality consistent with the corresponding semi-classical ICS phase.
\end{document}